\documentclass[aps,floatfix,superscriptaddress,showpacs,showkeys]{revtex4}
\usepackage{amssymb,amsmath,amsfonts,latexsym,graphicx,epsfig}
    
\usepackage{titlesec}
\usepackage{color}
\titleformat{\section}{\large\bfseries}{\thesection}{1em}{}

\newcommand{\bea}{\begin{eqnarray}}
\newcommand{\ena}{\end{eqnarray}}
\newcommand{\nn}{\nonumber\\}

\newcommand{\be}{\begin{equation}}
\newcommand{\en}{\end{equation}}

\newcommand{\la}{\langle}
\newcommand{\ra}{\rangle}
\newcommand{\Bla}{\Big\langle}
\newcommand{\Bra}{\Big\rangle}

\newcommand{\Tr}{\mbox{\rm{tr}}}

\begin{document}

\hfill DSF-2013-7 (Napoli), MITP/13-052 (Mainz) 

\title{
Polarization effects in the cascade decay 
\boldmath{$\Lambda_b \to  \Lambda(\to p\pi^-) + J/\psi(\to\ell^+\ell^-)$} \\
in the covariant confined quark model} 

\author{Thomas Gutsche}
\affiliation{Institut f\"ur Theoretische Physik, Universit\"at T\"ubingen,\\
Kepler Center for Astro and Particle Physics,\\
Auf der Morgenstelle 14, D-72076, T\"ubingen, Germany}

\author{Mikhail A. Ivanov}
\affiliation{Bogoliubov Laboratory of Theoretical Physics, \\
Joint Institute for Nuclear Research, 141980 Dubna, Russia}

\author{J\"{u}rgen G. K\"{o}rner}
\affiliation{PRISMA Cluster of Excellence, Institut f\"{u}r Physik, 
Johannes Gutenberg-Universit\"{a}t, \\ 
D-55099 Mainz, Germany}

\author{Valery E. Lyubovitskij}
\affiliation{Institut f\"ur Theoretische Physik, Universit\"at T\"ubingen,\\
Kepler Center for Astro and Particle Physics,\\
Auf der Morgenstelle 14, D-72076, T\"ubingen, Germany}
\affiliation{Department of Physics, Tomsk State University,
634050 Tomsk, Russia}

\author{Pietro Santorelli}
\affiliation{Dipartimento di Fisica, Universit\`a di Napoli
Federico II, Complesso Universitario di Monte S. Angelo,
Via Cintia, Edificio 6, 80126 Napoli, Italy}
\affiliation{Istituto Nazionale di Fisica Nucleare, Sezione di
Napoli, 80126 Napoli, Italy}

\today

\begin{abstract}

We calculate the invariant and helicity amplitudes for the nonleptonic
decay $\Lambda_b \to \Lambda\,+J/\psi,\psi(2S)$ in the covariant 
confined quark model. We discuss joint angular decay distributions 
in the cascade decay $\Lambda_b \to  \Lambda(\to p\pi^-) 
+J/\psi,\psi(2S) (\to\ell^+\ell^-)$ and calculate some of the asymmetry 
parameters that characterize the joint angular decay distribution. 
We confirm expectations 
from the naive quark model that the transitions into 
the $\lambda_{\Lambda}=1/2$ helicity states of the daughter baryon 
$\Lambda$ are strongly suppressed leading to a near maximal negative 
polarization of the $\Lambda$. For the same reason the azimuthal 
correlation between the two decay planes spanned by $(p\pi^{-})$ and 
$(\ell^{+}\ell^{-})$ is negligibly small. We provide form factor results 
for the whole accessible $q^{2}$ range. Our results are close to lattice 
results at minimum recoil and light-cone sum rule results at maximum recoil. 
A new feature of our analysis is that we include lepton mass effects in the 
calculation, which allows us to also describe the cascade decay 
$\Lambda_b \to  \Lambda(\to p\pi^-) + \psi(2S)(\to\tau^+\tau^-)$.

\end{abstract}


\maketitle

\newpage

\section{Introduction}

Recently the LHCb Collaboration has performed an angular 
analysis of the decay $\Lambda_b \to \Lambda + J/\psi$ 
where the $\Lambda_{b}$'s are produced in $pp$ collisions  
at $\sqrt{s} = 7$ TeV at the LHC (CERN)~\cite{Aaij:2013oxa}.  
They reported on the measurement of the relative magnitude of the helicity 
amplitudes in the decay $\Lambda_b \to \Lambda + J/\psi$ by a fit to
several asymmetry parameters in the cascade decay distribution
$\Lambda_b \to \Lambda(\to p\pi^-) + J/\psi (\to \ell^{+}\ell^{-})$. 
In the fit they were also able to measure the transverse polarization 
of the $\Lambda_b$ relative to the production plane.     
From a theoretical point of view the nonleptonic decay 
$\Lambda_b \to \Lambda + J/\psi$ 
is quite attractive in as much as the factorizable tree diagram is the only 
contribution to the decay; i.e. there are no $W$--exchange contributions 
[color compensation (C), 
exchange (E) and bow-tie (B) in the terminology of~\cite{Leibovich:2003tw}]  
as e.g. in  
$\Lambda_b \to \Lambda + \rho^{0}$. There have been a number of theoretical 
quark model calculations for the decay $\Lambda_b \to \Lambda + J/\psi$ 
that are based on the factorization 
hypothesis~\cite{Cheng:1995fe}-\cite{Mott:2011cx}. The 
results of some of these calculations have been
compared to the new experimental results by the LHCb Collaboration. 
We mention that the LHCb Collaboration has not given a result on the 
branching fraction $B(\Lambda_b \to \Lambda + J/\psi)$ for which the 
PDG quotes an average value of $(5.8 \pm 0.8) \times 10^{-4}$~\cite{pdg12}. 
The latter was deduced from the measurements by the CDF~\cite{Abe:1996tr}
and D0 Collaborations~\cite{Abazov:2011wt}. 

In this paper we present a detailed analysis of the decay process
$\Lambda_b \to \Lambda + J/\psi$ in the 
framework of the covariant quark model proposed and developed 
in Refs.~\cite{Ivanov:1996pz}-\cite{Branz:2010pq} for the study of 
mesons and baryons that are treated as bound 
states of their constituent quarks. Particle transitions are calculated 
from multiloop Feynman diagrams in which freely propagating constituent
quark fields connect the different nonlocal particle-quark vertices. We 
mention that the covariant quark model has recently been also applied to 
exotic tetraquark states \cite{Dubnicka:2010kz,Dubnicka:2011mm} and their 
decays. Quark confinement has been incorporated into the covariant quark 
model in an effective 
way ~\cite{Branz:2009cd}-\cite{Gutsche:2013pp} through an infrared 
regularization of the relevant quark-loop diagrams that removes quark 
thresholds in the loop diagrams (see details in 
Refs.~\cite{Branz:2009cd}-\cite{Gutsche:2013pp}). 

Our paper is structured as follows. 
In Sec.~II, we review the phenomenological aspects of 
the decay $\Lambda_b \to \Lambda\, + \, V$ 
where $V=J/\psi$ or $\psi(2S)$. This includes a discussion of 
kinematics, matrix elements, and invariant and helicity amplitudes. 
In Sec.~III
we write down joint angular decay distributions for the cascade decay 
$\Lambda_b \to \Lambda(\to p\pi^-) \,+ \,V\, (\to \ell^{+}\ell^{-})$ where
$V=J/\psi$ or $\psi(2S)$. We also define some pertinent decay asymmetry 
parameters that characterize the angular decay distributions. 
In Sec.~IV we review the salient features of the covariant confined quark 
model and present our form factor results, which we compare with the results
of other model calculations. In Sec.~V we carefully discuss the 
heavy quark limit (HQL) of
our $\Lambda_{b} \to \Lambda$ form factor expressions. 
In Sec.~VI we present our numerical results on helicity amplitudes, 
on the rate and on the asymmetry parameters in the decay processes   
$\Lambda_b \to \Lambda + J/\psi$ and 
$\Lambda_b \to \Lambda + \psi(2S)$. We have included the latter decay since it
allows us to discuss nonzero lepton mass effects in the kinematically allowed
decay
$\Lambda_b \to \Lambda + \psi(2S)\,(\to\tau^{+}\tau^{-})$. 
Finally, in Sec.~VII, we summarize our results. 

\section{\boldmath{$\Lambda_b \to \Lambda + J/\psi$} decay: 
matrix element  and observables} 

The effective Lagrangian~\cite{Altmannshofer:2008dz} for the 
$b\to s c \bar c$ transition is given by 
\be 
{\cal L}_{\rm eff} = \frac{G_F}{\sqrt{2}} \, 
V_{cb} \, V^\ast_{cs} \, \sum\limits_{i=1}^6 \, C_i \, Q_i \,, 
\label{eq:weak_Lag}
\en
where the $Q_i$ are the set of effective four-quark flavor-changing $b \to s$
operators 
\bea 
Q_1 &=& (\bar c^{a_1} O^\mu b^{a_2}) \, (\bar s^{a_2} O_\mu c^{a_1}) \,, 
\qquad
Q_4 = (\bar s^{a_1} O^\mu b^{a_2}) \, (\bar c^{a_2} O_\mu c^{a_1}) \,, 
\nn
Q_2 &=& (\bar c^{a_1} O^\mu b^{a_1}) \, (\bar s^{a_2} O_\mu c^{a_2}) \,, 
\qquad
Q_5 = (\bar s^{a_1} O^\mu b^{a_1}) \, (\bar c^{a_2} \tilde O_\mu c^{a_2}) \,, 
\nn
Q_3 &=& (\bar s^{a_1} O^\mu b^{a_1}) \, (\bar c^{a_2} O_\mu c^{a_2}) \,, 
\qquad 
Q_6 = (\bar s^{a_1} O^\mu b^{a_2}) \, (\bar c^{a_2} \tilde O_\mu c^{a_1}) \,, 
\label{eq:Q_i} 
\ena  
and where $V_{cb} = 0.0406$ and $V^\ast_{cs} = 0.974642$ are 
Cabibbo-Kabayashi-Maskawa (CKM) matrix elements; \\ 
$O^\mu = \gamma^\mu (1 - \gamma^5)$ and 
$\tilde O^\mu = \gamma^\mu (1 + \gamma^5)$. The 
$C_i$ are the set of Wilson coefficients~\cite{Altmannshofer:2008dz}: 
\be 
   C_1 = - 0.257\,, \quad 
   C_2 = 1.009\,,   \quad 
   C_3 = - 0.005\,, \quad
   C_4 = - 0.078\,, \quad 
   C_5 \simeq  0\,, \quad 
   C_6 = 0.001 \,. 
\label{eq:Wilson}
\en  
The quark-level matrix element contributing 
to the $\Lambda_b \to \Lambda + J/\psi$ decay
is given by 
\be
M(b\to s c \bar c)  = \frac{G_F}{\sqrt{2}} \,
C_{\rm eff}\, V_{cb} \, V^\ast_{cs} \, 
\left( \bar s \, O^\mu \, \,b \right)  \, 
\left( \bar c  \gamma_\mu c \right) \,, 
\label{eq:free_quark}
\en
where 
\be 
C_{\rm eff}= C_1 + C_3 + C_5 
+ \xi \Big(C_2 + C_4 + C_6\Big) \,.
\label{eq:color-factor}
\en 
The  color factor $\xi=1/N_c$
will be set to zero such that we keep only the leading term in the  
$1/N_c-$expansion.
The corresponding matrix elements of the exclusive transition
$\Lambda_b\to \Lambda\,+\,V$ is defined by  
\be
M(\Lambda_b\to \Lambda + V) =  
\frac{G_F}{\sqrt{2}} \, V_{cb} \, V^\ast_{cs} \, C_{\rm eff} \, 
f_V \, M_V \, \la \Lambda | \bar s O_\mu b | \Lambda_b \ra \, 
\epsilon^{\dagger\,\mu}(\lambda_V) \,,
\label{eq:matr_LbLJ}
\en 
where $M_V$ and $f_V$ are the mass and leptonic decay constant of $J/\psi$ 
or $\psi(2S)$. Note that the effective current 
$(\bar s \,O_\mu b)$ appearing in the set of operators in Eq.~(\ref{eq:Q_i})
is left-chiral. In the naive quark model where the spin of the 
$\Lambda_{b}$ and the $\Lambda$ are carried by the $b$ and $s$ quarks,
respectively, 
one would conclude that the $\Lambda$ is left-chiral and therefore
emerges with a dominant helicity of $\lambda_{\Lambda}=-1/2$. The
dominance of the $\lambda_{\Lambda}=-1/2$ helicity configuration predicted in 
the naive quark model is borne out by our exact calculation.

The hadronic matrix element $\la \Lambda | \bar s O_\mu b | \Lambda_b \ra$ 
in~(\ref{eq:matr_LbLJ}) is expanded in terms of dimensionless invariant 
form factors $f_i^{J}$ ($i=1, 2, 3$ and $J = V, A$), viz. 
\bea
\label{ffexpansion}
\la B_2\,|\,\bar s\, \gamma^\mu\, b\,| B_1 \ra &=&
\bar u_2(p_2)
\Big[ f^V_1(q^2) \gamma^\mu - f^V_2(q^2) i\sigma^{\mu q}/M_1
     + f^V_3(q^2) q^\mu/M_1 \Big] u_1(p_1)\,,
\nn
\la B_2\,|\,\bar s\, \gamma^\mu\gamma^5\, b\,| B_1 \ra &=&
\bar u_2(p_2)
\Big[ f^A_1(q^2) \gamma^\mu - f^A_2(q^2) i\sigma^{\mu q}/M_1
     + f^A_3(q^2) q^\mu/M_1 \Big]\gamma^5 u_1(p_1)\,, 
\label{eq:ff_def} 
\ena 
where $q = p_1 - p_2$. 
We have kept the scalar form factors $f_{3}^{V}$ and $f_{3}^{A}$ in the
form factor expansion Eq.~(\ref{ffexpansion}) although they do not contribute
to the decay $\Lambda_b \to \Lambda + J/\psi$ since
$q_\mu \, \epsilon_V^\mu  = 0$. The reason is that we want to compare our
results on the scalar form factor with the results of other
model calculations. The scalar form factors
would e.g. contribute to the rare decays 
$\Lambda_b \to \Lambda + \ell^{+}\ell^{-}$ and the decays 
$\Lambda_b \to \Lambda + \eta_{c}$ and 
$\Lambda_b \to p + \pi^{-}$.
The relevant form factors have been calculated before by us
in the covariant confined quark model~\cite{Gutsche:2013pp}. We shall
use the results of~\cite{Gutsche:2013pp} but we will add a few explanatory 
remarks concerning the cascade decay process
$\Lambda_b \to \Lambda(\to p\pi^-)\,+\,V(\to\ell^+\ell^-)$. 
We shall also present a detailed discussion of the HQL of our form factor 
expressions, which was not included in~\cite{Gutsche:2013pp}.  

As is well known it is convenient to analyze the 
decay in terms of helicity amplitudes 
$H_{\lambda_2\lambda_V}$ that are linearly
related to the invariant form factors $f_i^V$ and $f_i^A$ (see details in  
Refs.~\cite{Kadeer:2005aq,Faessler:2009xn,Branz:2010pq,Gutsche:2013pp}). 
Here we shall employ a generic notation such that the parent and daughter
baryons are denoted by $B_{1}$ and $B_{2}$.
The helicities of the daughter baryon $B_{2}$ and the vector charmonium state
$V$ are denoted by $\lambda_2$ and $\lambda_V$.
The pertinent relation is 
\be
H_{\lambda_2\lambda_V} =
\la \Lambda(\lambda_2) | \bar s O_\mu b | \Lambda_b(\lambda_1) \ra \, 
\epsilon^{\dagger\,\mu}(\lambda_V)
= H^{V}_{\lambda_2\lambda_V} - H^{A}_{\lambda_2\lambda_V}\,. 
\label{eq:hel_def}
\en 
The helicity amplitudes have been split into their  
vector $(H^V_{\lambda_2\lambda_V})$ 
and axial--vector $(H^A_{\lambda_2\lambda_V})$ parts. 
We shall work in the rest frame of 
the parent baryon $B_1$ with the daughter baryon $B_2$ moving in the
negative $z$ direction such that
$p_1^\mu = (M_1, {\bf 0})$, $p_2^\mu = (E_2, 0, 0,- |{\bf p}_2|)$ and
$q^\mu = (q_0, 0, 0,  |{\bf p}_2|)$. 
Further $q_0 = (M_+ M_- + q^2)/(2 M_1)$, 
$|{\bf p}_2| = \sqrt{Q_+Q_-}/{2 M_1}$  
and 
$E_2 = M_1 - q_0 = (M_1^2 + M_2^2 - q^2)/(2 M_1)$, where $q^2 = M_V^2$ 
for the on-mass shell $J/\psi(\psi(2S))$ meson. 
We have introduced the notation 
$M_\pm = M_1 \pm M_2$, $Q_\pm = M_\pm^2 - q^2$. 
Angular momentum conservation fixes the helicity $\lambda_1$
of the parent baryon such that $\lambda_1 = - \lambda_2 + \lambda_V$. 
The relations between the helicity amplitudes $H^{V,A}_{\lambda_2\lambda_V}$ 
and the invariant amplitudes are given 
by~\cite{Gutsche:2013pp} 
\bea 
H^V_{\pm\frac{1}{2} \pm 1} &=& \sqrt{2 Q_-} \, 
\biggl( f_1^V + \frac{M_+}{M_1} \, f_2^V \biggr)\,, 
\qquad
H^A_{\pm\frac{1}{2} \pm 1} = \pm \sqrt{2 Q_+} \, 
\biggl( f_1^A - \frac{M_-}{M_1} \, f_2^A \biggr)\,,
\nn
H^V_{\pm\frac{1}{2}  0} &=& \sqrt{\frac{Q_-}{q^2}} \,  
\biggl( M_+ \, f_1^V + \frac{q^2}{M_1} \, f_2^V \biggr)\,, 
\qquad
H^A_{\pm\frac{1}{2} 0} = \pm\sqrt{\frac{Q_+}{q^2}} \,  
\biggl( M_- \, f_1^A  - \frac{q^2}{M_1} \, f_2^A \biggr)\,. 
\label{eq:hel_ff} 
\ena 
As in Ref.~\cite{Gutsche:2013pp} we  introduce the following 
combinations of helicity amplitudes:  
\bea 
\begin{array}{lr}
\mbox{$ H_U = |H_{\frac{1}{2}1}|^2 +  |H_{-\frac{1}{2}-1}|^2$} &
\hfill\mbox{ \rm transverse unpolarized}\,, 
\\
\mbox{$ H_L = |H_{\frac{1}{2}0}|^2 +  |H_{-\frac{1}{2}0}|^2$} &
\hfill\mbox{ \rm longitudinal unpolarized}\,. 
\\
\end{array}
\label{eq:hel_comb}
\ena   
The partial helicity width corresponding to the two specific combinations
of helicity amplitudes in~(\ref{eq:hel_comb}) is defined by 
($\varepsilon=m_l^2/M^2_{V};\, 
v^{2}=1-4\varepsilon$) 
\be
\Gamma_{I}(\Lambda_b \to \Lambda\,+\,V) 
= \frac{G_F^2}{32 \pi} \, \frac{|{\bf p}_2|}{M_1^2} \, 
|V_{cb} V^\ast_{cs}|^2 \, C_{\rm eff}^2 \, f_V^2 \, M_V^2 
v(1+2\varepsilon)
H_I
 \quad\quad I = U, L \,.  
\label{eq:hel_partial_width}
\en 
For the $\Lambda_b \to \Lambda\,+\,V$ decay width one finds 
\bea
\Gamma(\Lambda_b \to \Lambda\,+\,V) = \Gamma_U + \Gamma_L \,. 
\label{eq:LbLJ-decay}
\ena 

%%%%%%%%%%%%%%%%%%%%%%%%%%%%%%%%%%%%%%%%%%%%%%%%%%%%%%%%%%%%%%%%%%%%%%%%%%%%%

\section{Joint angular decay distributions in the cascade decay \\ 
\boldmath{$\Lambda_b\to \Lambda(\to p\pi^-)+V(\to\ell^+\ell^-)$}}

%%%%%%%%%%%%%%%%%%%%%%%%%%%%%%%%%%%%%%%%%%%%%%%%%%%%%%%%%%%%%%%%%%%%%%%%%%%%

As in the case of the rare meson decays $B \to K^{(\ast)} + \ell^+\ell^-$
 $(\ell=e,\mu,\tau)$ treated in~\cite{Faessler:2002ut} one 
can exploit the cascade nature of the decay
$\Lambda_{b}(\uparrow) \to \Lambda (\to p \pi^{-})+ 
V(\to \ell^{+}\ell^{-})$ of polarized $\Lambda_{b}(\uparrow)$ decays to
write down a fivefold angular decay distribution involving the polar angles 
$\theta_{1},\, \theta_{2}$ and $\theta$, and the two azimuthal angles
$\phi_{1}$ and $\phi_{2}$. 
$V$ stands for $J/\psi$ or $\psi(2S)$. Since the decay 
$\psi(2S)\to \tau^{+}\tau^{-}$ is kinematically allowed, we include lepton mass
effects in our decay formulas. 
The angular decay distribution involves
the helicity amplitudes $h^{V}_{\lambda_{\ell^{+}}\lambda_{\ell^{-}}}$ for the 
decay $V \to \ell^{+}\ell^{-}$,
$H_{\lambda_{\Lambda}\lambda_{V}}$ for the decay 
$\Lambda_{b} \to \Lambda + V$ and $h^{B}_{\lambda_{p}0}$ 
for the decay $\Lambda \to p + \pi^{-}$. 

We do not write out the full fivefold angular decay distribution that can 
be found in~\cite{lednicky86}, or that can be adapted from the
corresponding fivefold decay distributions for the semileptonic baryon
decays $\Xi^{0}\to \Sigma^{+}+\ell^{-}\bar \nu_{\ell}$ and 
$\Lambda_{c}\to \Lambda+\ell^{+} \nu_{\ell}$ 
written down in ~\cite{Kadeer:2005aq} and \cite{Bialas:1992ny}, respectively. 
Instead we discuss 
a threefold polar angle distribution for polarized $\Lambda_{b}$ decay
and a threefold joint decay distribution for unpolarized $\Lambda_{b}$ decay.
These can be obtained from the full fivefold decay distributions written down
in~\cite{Kadeer:2005aq,Bialas:1992ny,lednicky86} by the 
appropriate angular integrations or by setting the polarization of the
$\Lambda_{b}$ to zero. 

%%%%%%%%%%%%%%%%%%%%%%%%%%%%%%%%%%%%%%%%%%%%%%%%%%%%%%%%%%%%%%%%%%%%%%%%%%%%

\subsection{
Polar angle distribution in polarized 
\boldmath{$\Lambda_b$} decay}

%%%%%%%%%%%%%%%%%%%%%%%%%%%%%%%%%%%%%%%%%%%%%%%%%%%%%%%%%%%%%%%%%%%%%%%%%%%%%

\begin{figure}[ht] 
\begin{center}
\epsfig{figure=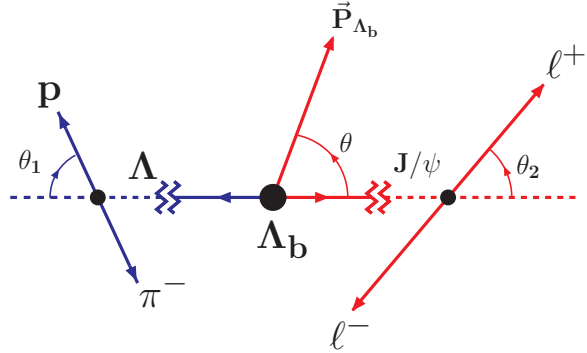,scale=.6} 
\caption{Definition of 
the three polar angles
in the cascade decay $\Lambda_{b}(\uparrow) \to \Lambda (\to p \pi^{-})
+ J/\psi(\to \ell^{+}\ell^{-})$ 
of a polarized $\Lambda_b$ baryon.} 
\label{fig:LbLJ_polar}
\end{center}
\end{figure}

Let us first consider the polar angle distribution 
$W(\theta,\theta_1,\theta_2)$ for polarized $\Lambda_{b}$ decays which has 
been discussed before in~\cite{Aaij:2013oxa,Hrivnac:1994jx} in the zero 
lepton mass approximation (see Fig.~\ref{fig:LbLJ_polar}). 

The angular decay distribution can be derived from the master formula 
\bea
W(\theta,\theta_1,\theta_2)\propto  
\tfrac12\sum_{\rm helicities}
|h^{V}_{\lambda_{\ell^{+}}\lambda_{\ell^{-}}}|^2
\left[d^1_{\lambda_V,\lambda_{\ell^{+}}-\lambda_{\ell^{-}}}(\theta_{2})
\right]^2
\rho_{\lambda_{\Lambda_b},\lambda_{\Lambda_b}}(\theta)
\delta_{\lambda_{\Lambda_b},\lambda_V-\lambda_\Lambda}
|H_{\lambda_{\Lambda}\lambda_{V}}|^2
\left[d^{1/2}_{\lambda_{\Lambda}\lambda_p}(\theta_1)\right]^2
|h^{B}_{\lambda_p 0}|^2\, , 
\label{eq: master}
\ena
where the summation extends over all possible helicities
$\lambda_{\ell^{+}},\lambda_{\ell^{-}},\lambda_{\Lambda_b},\lambda_\Lambda,
\lambda_p=\pm \tfrac12$ and $\lambda_V=0,\pm 1$. 
For the diagonal terms of the density matrix 
$\rho_{\lambda_{\Lambda_b},\lambda_{\Lambda_b}}(\theta)$ appearing in 
Eq.~(\ref{eq: master})
one has 
\begin{eqnarray}
\rho(\theta)=\frac12 {\rm diag}(1-P_b\cos\theta,1+P_b\cos\theta)\,. 
\label{eq:dens-matr}
\end{eqnarray} 
The vector current lepton helicity amplitudes are given by 
(see \cite{Gutsche:2013pp})
\be
\mbox{flip:}\quad h^{V}_{-\tfrac12 -\tfrac12} = 
h^{V}_{+\tfrac12 +\tfrac12} =2m_l\,, \qquad \mbox {nonflip:}\quad
h^{V}_{-\tfrac12 +\tfrac12} = h^{V}_{+\tfrac12 -\tfrac12} =\sqrt{ 2 q^2}\,.
\label{flipnoflip}
\en
Finally, factorizing out the combination 
$|h^{B}_{+\tfrac12 0}|^2+|h^{B}_{-\tfrac12 0}|^2\propto 
{\rm Br}(\Lambda\to p \pi)$ and introducing the asymmetry parameter 
\be
\alpha_\Lambda = \frac{|h^{B}_{+\tfrac12 0}|^2 - |h^{B}_{-\tfrac12 0}|^2}
                    {|h^{B}_{+\tfrac12 0}|^2 + |h^{B}_{-\tfrac12 0}|^2}\,,
\label{eq:asym_lambda}
\en
one obtains the angular decay distribution 

\bea
W(\theta,\theta_1,\theta_2)  
&\propto&
\frac12 |H_{+\tfrac12 1}|^2
\left[ q^2(1+\cos^2\theta_{2}) + 4 m_l^2 \sin^2\theta_{2} \right]
(1-P\cos\theta)(1+\alpha_\Lambda\cos\theta_{1})
\nn
&+&
\frac12 |H_{-\tfrac12 -1}|^2
\left[ q^2(1+\cos^2\theta_{2}) + 4 m_l^2 \sin^2\theta_{2} \right]
(1+P\cos\theta)(1-\alpha_\Lambda\cos\theta_{1})
\nn
&+&
|H_{+\tfrac12 0}|^2\left[ q^2 \sin^2\theta_{2} 
+ 4 m_l^2 \cos^2\theta_{2} \right]
(1+P\cos\theta)(1+\alpha_\Lambda\cos\theta_{1})
\nn
&+&
|H_{-\tfrac12 0}|^2\left( q^2 \sin^2\theta_{2} 
+ 4 m_l^2 \cos^2\theta_{2} \right)
(1-P\cos\theta)(1-\alpha_\Lambda\cos\theta_{1})\,.
\label{eq:full-dis}
\ena
Following Ref.~\cite{Aaij:2013oxa} we introduce linear combinations of
normalized squared helicity amplitudes 
$ |\widehat H_{\lambda_{\Lambda_{b}} \lambda_{V}}|^2$ by writing
\bea
\alpha_b &=& 
  |\widehat H_{+\tfrac12 0}|^2 - |\widehat H_{-\tfrac12 0}|^2 
+ |\widehat H_{-\tfrac12 -1}|^2 - |\widehat H_{+\tfrac12 +1}|^2
\,,\nn
r_0 &=& |\widehat H_{+\tfrac12 0}|^2 + |\widehat H_{-\tfrac12 0}|^2 
\,,\nn    
r_1 &=& |\widehat H_{+\tfrac12 0}|^2 - |\widehat H_{-\tfrac12 0}|^2 
\,,
\label{eq:asym-param}
\ena
where $ |\widehat H_{\lambda_{\Lambda_{b}} \lambda_{V}}|^2=
|H_{\lambda_{\Lambda_{b}} \lambda_{V}}|^2/N$ 
and where the normalization factor $N$ is given by
$N\equiv |H_{+\tfrac12 0}|^2 + |H_{-\tfrac12 0}|^2
        + |H_{-\tfrac12 -1}|^2 + |H_{+\tfrac12 +1}|^2$. 
Similar to~\cite{Aaij:2013oxa} the angular decay distribution can be 
rearranged into the form 
\bea
\label{eq:w3}
\widetilde{W}(\theta,\theta_1,\theta_2)  
&=& \sum_{i=0}^7 \; 
f_i(\alpha_b,r_0,r_1)\;
g_i(P_b,\alpha_\Lambda)\;
h_i(\cos\theta,\cos\theta_1,\cos\theta_2)\; 
\ell_i(\varepsilon) \nonumber\\
&=& v\,(1+2\varepsilon)+\sum_{i=1}^7 \; 
f_i(\alpha_b,r_0,r_1)\;
g_i(P_b,\alpha_\Lambda)\;
h_i(\cos\theta,\cos\theta_1,\cos\theta_2)\; 
\ell_i(\varepsilon)\,, 
\ena
such that the angular factors $h_i(\cos\theta,\cos\theta_1,\cos\theta_2)$
$(i=1,\ldots,7)$ in the second row of Eq.~(\ref{eq:w3}) 
integrate to zero after polar integration.
The functions $f_{i},\,g_{i},\,h_{i}$ and $\ell_{i}$ $(i=0,\ldots,7)$ 
that describe the normalized angular distribution~(\ref{eq:w3}) are listed 
in Table~\ref{tab:angdistrb3dim}. Setting $\varepsilon=m^{2}_{\ell}/M_{V}^{2}$
to zero and the velocity parameter 
${v}=\sqrt{1-4\varepsilon}$ to $1$ as is appropriate in the
zero lepton mass approximation one recovers Table~\ref{tab:velocity} of 
Ref.~\cite{Aaij:2013oxa}.
It is clear that one can determine the four 
parameters ($P_b$, $\alpha_b$, $r_0$ and $r_1$) from a global fit to the 
polar angle distribution as has been done in~\cite{Aaij:2013oxa}. 

\begin{table}[htb]
  \caption{
Decay functions appearing in the threefold polar angle 
distribution in the decay of a polarized $\Lambda_{b}$. \\
The velocity is defined as $v=\sqrt{1-4\varepsilon}.$
In our numerical analysis we use $\alpha_{\Lambda}=0.642$~\cite{pdg12}.}
  \begin{center}
    \begin{tabular}{llllc}
\hline\noalign{\vskip 2mm}
$i$ \hspace*{0.5cm} & $f_i(\alpha_b,r_0,r_1)$  \hspace*{0.5cm} & 
$g_i(P_b,\alpha_\Lambda)$  \hspace*{0.5cm} & 
    $h_i(\cos\theta,\cos\theta_1,\cos\theta_2)$ & $\ell_i(\varepsilon) $
\\ \noalign{\vskip 2mm}\hline\noalign{\vskip 2mm}
0   & $1$   & $1$    & $1$ &   $v\cdot(1+2\varepsilon)$
\\ \noalign{\vskip 1mm}
1   & $\alpha_b$  & $P_b$  & $\cos\theta$  &   
    $\mbox{\em v}\cdot(1+2\varepsilon)$
\\ \noalign{\vskip 1mm}
2   & $2 r_1-\alpha_b$  & $\alpha_\Lambda$  & $\cos\theta_1$  &   
      $\mbox{\em v}\cdot(1+2\varepsilon)$
\\ \noalign{\vskip 1mm}
3   & $2 r_0-1$           & $ P_b\alpha_\Lambda$  
& $\cos\theta\cos\theta_1$  & 
      $\mbox{\em v}\cdot(1+2\varepsilon)$
\\ \noalign{\vskip 1mm}
4   & $\tfrac12(1-3 r_0)$ & $1$   & $\tfrac12(3\cos^2\theta_2-1)$   &   
      $\mbox{\em v}\,\cdot\mbox{\em v}^{2}$  
\\ \noalign{\vskip 1mm}
5   & $\tfrac12(\alpha_b-3 r_1)$  & $P_b$  &   
      $\tfrac12(3\cos^2\theta_2-1)\cos\theta$   &   
      $\mbox{\em v}\,\cdot\mbox{\em v}^{2}$ 
\\ \noalign{\vskip 1mm}
6   & $-\tfrac12 (\alpha_b + r_1)$  & $\alpha_\Lambda$  &  
      $\tfrac12(3\cos^2\theta_2-1)\cos\theta_1$  &   
      $\mbox{\em v}\,\cdot\mbox{\em v}^{2}$      
\\ \noalign{\vskip 1mm}
7   & $-\tfrac12 (1 + r_0)$    & $P_b\alpha_\Lambda$  &  
      $\tfrac12(3\cos^2\theta_2-1)\cos\theta\cos\theta_1$ &
      $\mbox{\em v}\,\cdot\mbox{\em v}^{2}$ 
\\[1.5ex]
\hline
    \end{tabular}
  \end{center}
  \label{tab:angdistrb3dim}
\end{table}

\begin{table}[hb]
\begin{center}
\caption{Numerical values of the velocity $v$ and the velocity
factors $\ell_{i}(\varepsilon)$.} 
\vspace*{.25cm}
\def\arraystretch{1}
    \begin{tabular}{|c|c|c|c|}
      \hline
& \,$J/\psi\to\mu^{+}\mu^{-}$\, &$\psi(2S)\to \mu^{+}\mu^{-}$ & 
$\psi(2S)\to \tau^{+}\tau^{-}$\\ \hline
$v$        &  0.998 & 0.998&0.266 \\
\hline
\,$v(1+2\varepsilon)$\, &  1.000 & 1.000 & 0.389 \\
$(i=0,\ldots,3)$  &        &       &      \\
\hline
$v^{3}$  &  0.993 & 0.995 &0.019 \\
$(i=4,\ldots,7)$  &        &       &      \\
\hline
\end{tabular}
\label{tab:velocity}
\end{center}
\end{table}

Let us briefly dwell on the powers of the velocity factor $v$ in 
Table~\ref{tab:angdistrb3dim}.
The common factor $v$ in the fifth column of 
Table~\ref{tab:angdistrb3dim} has its origin in the
phase space factor ${v}$ in the decay $V\to\ell^{+}\ell^{-}$. The
remaining ${v}$ dependence results from a dominant $S$--wave 
contribution in the factor $(1+2\varepsilon)$ and 
a dominant $S-D$--interference contribution in the factor 
${v}^{2}=(1-4\varepsilon)$, 
respectively, as can be seen by performing an $LS$ analysis of
the decay $V\to\ell^{+}\ell^{-}$. The $LS$ amplitudes $M_{LS}$ are given by 
$M_{01}=\sqrt{2/3}\,(h^{V}_{+\tfrac12 +\tfrac12}+
\sqrt{2}h^{V}_{+\tfrac12 -\tfrac12})$
and
$M_{21}=\sqrt{2/3}\,(-\sqrt{2}h^{V}_{+\tfrac12 +\tfrac12}+
h^{V}_{+\tfrac12 -\tfrac12})$. One then finds
\bea 
\label{parwave}
\phantom{{v}^{2}=}1+2\varepsilon&=&\frac{1}{4q^{2}}(M_{01}^{2}+M_{21}^{2})\,, 
\nonumber \\
{v}^{2}=1-4\varepsilon&=&\frac{1}{4q^{2}}M_{21}(2\sqrt{2}M_{01}-M_{21}) \,. 
\ena

By integrating over two respective angles of the three polar angles one 
obtains the single angle 
distributions
$W(\theta)$, $W(\theta_{1})$ 
and $W(\theta_{2})$. 
In their normalized forms they read
\bea
\widehat{W}(\theta)&=&
\tfrac{1}{2}\,
\Big(
1+\alpha_{b}P_{b}\cos\theta
\Big)\,,
\label{wtheta} \\
\widehat{W}(\theta_{1})&=&
\tfrac{1}{2}\,
\Big(
1+2(2r_{1}-\alpha_{b})\alpha_{\Lambda}\cos\theta_{1}
\Big)\,,
\label{wtheta1} \\
\widehat{W}(\theta_{2})&=&
\frac{1}{2\,(1+2\varepsilon)}\,
\Big(
(1+2\varepsilon)+\tfrac14(1-4\varepsilon)\,( 1-3r_{0})(3\cos^2\theta_{2}-1)
\Big)\,,
\label{wtheta2}
\end{eqnarray}
where the factor 
\bea 
(2r_{1}-\alpha_{b})= 
  |\widehat H_{+\tfrac12 0}|^2 
+ |\widehat H_{+\tfrac12 1}|^2 
- |\widehat H_{-\tfrac12 0}|^2 
- |\widehat H_{-\tfrac12 -1}|^2 = P^{\ell}_{\Lambda} 
\ena 
defines the longitudinal polarization of the daughter baryon $\Lambda$ 
and the factor 
\bea 
(1-3r_{0})=-2 (
  |\widehat H_{+\tfrac12 0}|^2 
+ |\widehat H_{-\tfrac12 0}|^2)
+ |\widehat H_{+\tfrac12 1}|^2 
+ |\widehat H_{-\tfrac12 -1}|^2 
\ena
is a measure of the longitudinal/transverse polarization composition 
of the vector charmonium state. 
Since the decay $V\to\ell^+\ell^-$ is electromagnetic 
and therefore parity conserving, the decay is not sensitive to the difference  
of the transverse-plus and transverse-minus helicity contributions
$|\widehat H_{+\tfrac12 +1}|^2 - |\widehat H_{-\tfrac12 -1}|^2$. For the
same reason there is no linear $\cos\theta_2$ contribution in 
Eq.~(\ref{wtheta2}). 

The distributions $\widehat{W}(\theta)$ and $\widehat{W}(\theta_{1})$ are 
asymmetric in $\cos\theta$ and $\cos\theta_{1}$ such that they can be
characterized by the forward-backward (FB) asymmetries
\bea
A_{FB}\,\big|_{\theta}&=& \tfrac12 \,\alpha_{b}P_{b}\,,
\label{afbtheta}\\
A_{FB}\,\big|_{\theta_{1}}&=& (2r_{1}-\alpha_{b})\alpha_{\Lambda}\,. 
\label{afbtheta1}
\ena
The distribution $\widehat{W}(\theta_{2})$ is symmetric in $\cos\theta_{2}$ 
with its convexity parameter $c_{f}$ given by 
\be
\label{convex}
c_{f}=\frac{d^{2}\widehat{W}(\cos\theta_{2})}{d(\cos\theta_{2})^{2}}=
\frac34\, \frac{(1-4\varepsilon)}{(1+2\varepsilon)}\,(1-3r_{0}) \,. 
\en

In Table~\ref{tab:velocity} we have listed the numerical values of the 
two velocity factors $v(1+2\varepsilon)$ and $v^{3}$ for the cases involving
the muon and the tau lepton. It is quite apparent that the zero lepton mass 
approximation is quite good for $J/\psi\to\mu^{+}\mu^{-}$ and 
$\psi(2S)\to \mu^{+}\mu^{-}$. For the case 
$\psi(2S)\to \tau^{+}\tau^{-}$ the factor $v(1+2\varepsilon)$ provides a
reduction of about $60\%$ relative to the $e^{+}e^{-}$ and $\mu^{+}\mu^{-}$
cases while the factor $v^{3}$ becomes negligibly small. This means
that for $\psi(2S)\to \tau^{+}\tau^{-}$ one loses the analyzing
power of the lepton--side decay, i.e. of the last four rows of 
Table~\ref{tab:angdistrb3dim}. For the same reason, one will have an almost 
flat decay distribution $\widehat{W}(\cos\theta_{2})$ for 
$\psi(2S)\to \tau^{+}\tau^{-}$ because of the factor $v^{2}=1-4\varepsilon$
in the expression Eq.~(\ref{convex}) for the convexity parameter.  

%%%%%%%%%%%%%%%%%%%%%%%%%%%%%%%%%%%%%%%%%%%%%%%%%%%%%%%%%%%%%%%%%%%%%%%%%%%%%%%

\subsection{Azimuthal angle distribution} 

%%%%%%%%%%%%%%%%%%%%%%%%%%%%%%%%%%%%%%%%%%%%%%%%%%%%%%%%%%%%%%%%%%%%%%%%%%%%%%

\begin{figure}[ht] 
\begin{center}
\epsfig{figure=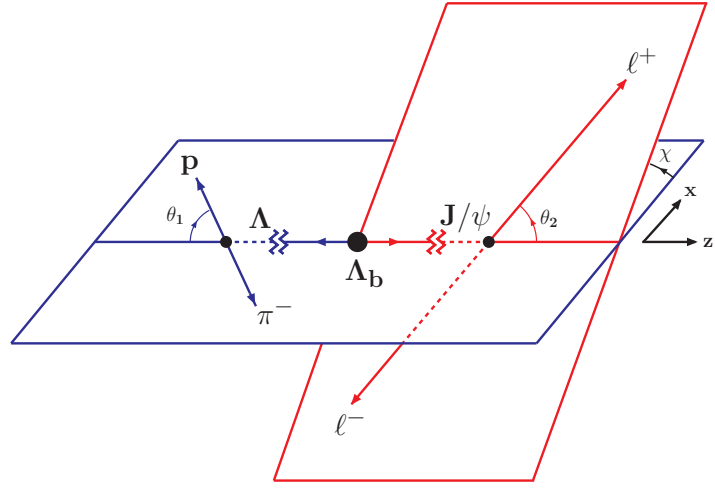,scale=.45} 
\caption{Definition of two polar angles $\theta_1$, $\theta_2$ and one 
azimuthal angle $\chi$ in the cascade decay
$\Lambda_{b} \to \Lambda (\to p \pi^{-})
+ J/\psi(\to \ell^{+}\ell^{-})$ of an unpolarized $\Lambda_b$.}

\label{fig:LbLJ_unpolar}
\end{center}
 \end{figure}

Consider the angular decay distribution of an unpolarized $\Lambda_{b}$ that 
is characterized by the two polar angles $(\theta_{1})$ and $(\theta_{2})$,
and an azimuthal angle $\chi$ defined by the azimuth of the two decay planes
defined by the decays $V\to\ell^{+}\ell^{-}$ and $\Lambda \to p\pi^{-}$ 
(see Fig.~\ref{fig:LbLJ_unpolar}).
The angular decay 
distribution can then be calculated from the master formula
\bea
W(\theta_1,\theta_2,\chi)  
&\propto& 
\sum_{\rm helicities}
|h^{V}_{\lambda_1\lambda_2}|^2\,e^{i(\lambda_V-\lambda'_V)\chi}
d^1_{\lambda_V,\lambda_1-\lambda_{2}}(\theta_{2})
d^1_{\lambda'_V,\lambda_1-\lambda_{2}}(\theta_{2})\nonumber\\
&\times&
\delta_{\lambda_V-\lambda_\Lambda,\lambda'_V-\lambda'_\Lambda}
H_{\lambda_{\Lambda}\lambda_{V}}
H^{\dagger}_{\lambda'_{\Lambda}\lambda'_{V}}
d^{1/2}_{\lambda_{\Lambda}\lambda_p}(\theta_1)
d^{1/2}_{\lambda'_{\Lambda}\lambda_p}(\theta_1)
|h^{B}_{\lambda_p 0}|^2 \,. 
\label{azimuth}
\ena
Let us first present a qualitative argument that the azimuthal correlation 
between the two decay planes is small. Azimuthal correlations result from the
configurations $\lambda_V-\lambda'_V=\pm1$. This implies that 
$\lambda_\Lambda= -\lambda'_\Lambda$ following from the $\delta$--function
condition $\lambda_V-\lambda_\Lambda=\lambda'_V-\lambda'_\Lambda$, which again
follows from the fact that
$\Lambda_{b}$ is treated as unpolarized. The azimuthal correlations
are therefore determined by bilinear forms such as $H_{\tfrac 12 \lambda_V}
H^{\dagger}_{-\tfrac 12 \lambda'_V}$ with $\lambda_V\neq\lambda'_V$. We shall 
see in Sec.~VI that in general 
$|H_{\tfrac 12 \lambda_V}|\ll|H_{-\tfrac 12 \lambda'_V}|$ (as also expected 
from the naive quark model) such that one concludes that the azimuthal 
correlations between the two decay planes are quite small.

Let us cast this reasoning into a more quantitative form. The threefold 
angular decay distribution resulting from Eq.~(\ref{azimuth}) reads

\begin{eqnarray}
W(\theta_{1},\theta_{2},\chi)&=& 
\frac 38 (1+\cos^{2}\theta_{2})\Big(|H_{\tfrac 12 1}|^{2}
(1+\alpha_{\Lambda}\cos\theta_{1}) 
+|H_{-\tfrac 12 -1}|^{2}
(1-\alpha_{\Lambda}\cos\theta_{1})\Big) \nonumber \\
&+&\frac 34\sin^{2}\theta_{2}\Big(|H_{\tfrac 12 0}|^{2}
(1+\alpha_{\Lambda}\cos\theta_{1}) 
+|H_{-\tfrac 12 0}|^{2}
(1-\alpha_{\Lambda}\cos\theta_{1})\Big) \nonumber \\
&+&\frac{4m^{2}_{\ell}}{q^{2}}\,\,\bigg[
\frac 38 \sin^{2}\theta_{2}\Big(|H_{\tfrac 12 1}|^{2}
(1+\alpha_{\Lambda}\cos\theta_{1}) 
+|H_{-\tfrac 12 -1}|^{2}
(1-\alpha_{\Lambda}\cos\theta_{1})\Big) \nonumber \\
&+&\frac 34\cos^{2}\theta_{2}\Big(|H_{\tfrac 12 0}|^{2}
(1+\alpha_{\Lambda}\cos\theta_{1}) 
+|H_{-\tfrac 12 0}|^{2}
(1-\alpha_{\Lambda}\cos\theta_{1})\Big)\bigg] \nonumber \\
&+&(1-\frac{4m^{2}_{\ell}}{q^{2}})\frac{3}{4\sqrt{2}}
\alpha_{\Lambda}\sin2\theta_{2}\sin\theta_{1}
\Big(\cos\chi\rm{Re}\big[ H_{\tfrac 12 1}H^{\dagger}_{-\tfrac 12 0}- 
H_{-\tfrac 12 -1}H^{\dagger}_{\tfrac 12 0}\big]\nonumber \\
&-&
\sin\chi{\rm{Im}}\big[ H_{\tfrac 12 1}H^{\dagger}_{-\tfrac 12 0}+ 
H_{-\tfrac 12 -1}H^{\dagger}_{\tfrac 12 0}\big]\Big) \,. 
\end{eqnarray}
Integrating over the hadron--side polar angle $\theta_{1}$ one obtains
\begin{eqnarray}
\label{azimuth2}
W(\theta_{2},\chi)&=& 
\frac 38 (1+\cos^{2}\theta_{2})\,\cdot\,2\Big(|H_{\tfrac 12 1}|^{2} 
+|H_{-\tfrac 12 -1}|^{2}\Big) 
+\frac 34\sin^{2}\theta_{2}\,\cdot\,2\Big(|H_{\tfrac 12 0}|^{2} 
+|H_{-\tfrac 12 0}|^{2}\Big) \nonumber \\
&+&\frac{4m^{2}_{\ell}}{q^{2}}\,\,\bigg[
\frac 38 \sin^{2}\theta_{2}\,\cdot\,2\Big(|H_{\tfrac 12 1}|^{2} 
+|H_{-\tfrac 12 -1}|^{2}\Big) 
+\frac 34\cos^{2}\theta_{2}\,\cdot\,2\Big(|H_{\tfrac 12 0}|^{2} 
+|H_{-\tfrac 12 0}|^{2}\Big)\bigg] \nonumber \\
&+&(1-\frac{4m^{2}_{\ell}}{q^{2}})\frac{3}{4\sqrt{2}}\frac{\pi}{2}
\alpha_{\Lambda}\sin2\theta_{2}
\Big(\cos\chi\,\rm{Re} \big(H_{\tfrac 12 1}H^{\dagger}_{-\tfrac 12 0}- 
 H_{-\tfrac 12 -1}H^{\dagger}_{\tfrac 12 0}\big)\nonumber \\
&-&\sin\chi\,{\rm{Im}} \big(H_{\tfrac 12 1}H^{\dagger}_{-\tfrac 12 0}+ 
H_{-\tfrac 12 -1}H^{\dagger}_{\tfrac 12 0}\big)\Big) \,. 
\end{eqnarray}
If one wants to define a measure of the azimuthal correlation, one cannot 
integrate Eq.~(\ref{azimuth2}) over the whole range of the 
lepton--side polar angle
$\theta_{2}$ because $\int_{-1}^{1}d\cos\theta_{2}\sin2\theta_{2}=0$. However,
one can recover a nonzero azimuthal correlation measure by defining a
FB asemmtry with respect to the lepton-side polar
angle $\theta_{2}$ by writing 
\begin{eqnarray}
A_{FB}(\chi) = \frac{F-B}{F+B} \,. 
\end{eqnarray}
On reintroducing the normalized helicity amplitudes one obtains
\begin{eqnarray}
\label{azi}
A_{FB}(\chi)= \frac{(1-4\varepsilon)}{(1+2\varepsilon)}
\frac{\pi}{8\sqrt{2}}\alpha_{\Lambda}
\biggl(\cos\chi\rm{Re}\biggl[\widehat H_{\tfrac 12 1}
\widehat H^{\dagger}_{-\tfrac 12 0}- 
\widehat H_{-\tfrac 12 -1}\widehat H^{\dagger}_{\tfrac 12 0}\biggr] 
\,-\,\sin\chi{\rm{Im}}\biggl[ \widehat H_{\tfrac 12 1}
\widehat H^{\dagger}_{-\tfrac 12 0}+ 
\widehat H_{-\tfrac 12 -1}\widehat H^{\dagger}_{\tfrac 12 0}\biggr]\biggr) 
\,.  
\end{eqnarray} 

%%%%%%%%%%%%%%%%%%%%%%%%%%%%%%%%%%%%%%%%%%%%%%%%%%%%%%%%%%%%%%%%%%%%%%%%%%%%%%

\section{The \boldmath{$\Lambda_b\to\Lambda$} form factors in the
covariant confined quark model} 

%%%%%%%%%%%%%%%%%%%%%%%%%%%%%%%%%%%%%%%%%%%%%%%%%%%%%%%%%%%%%%%%%%%%%%%%%%%%%%

For the description of the couplings of the baryons $\Lambda_Q$ ($Q=b,s$)
and the charmonium vector meson states $V = J/\psi, \psi(2S)$ to their 
three and two constituent 
quarks, respectively,
we employ a generic Lagrangian that reads  
\bea
\Lambda_Q:\qquad &&{\cal L}^{\Lambda_Q}_{\rm int}(x) 
 = g_{\Lambda_Q} \,\bar \Lambda_Q(x)\cdot J_{\Lambda_Q}(x) + \mathrm{H.c.}\,, 
\label{eq:lag_Lambda}\\
\phantom{\Lambda_Q:}\qquad &&J_{\Lambda_Q}(x) 
= \int\!\! dx_1 \!\! \int\!\! dx_2 \!\! \int\!\! dx_3 \, 
F_{\Lambda_Q}(x;x_1,x_2,x_3) \, 
\epsilon^{a_1a_2a_3} \, Q^{a_1}(x_1)\,u^{a_2}(x_2) \,C \, \gamma^5 \, 
d^{a_3}(x_3)\,,
\nn
V:\qquad &&{\cal L}^{V}_{\rm int}(x) 
= g_{V} \, V(x)\cdot J_{V}(x)\,,  
\label{eq:lag_psi}\\
\phantom{J/\psi:} \qquad && J_{V}(x) = \int\!\! dx_1 \!\! \int\!\! dx_2 \, 
F_{V}(x;x_1,x_2) \, \bar c^{a}(x_1) \,\gamma^\mu c^a(x_2) \,. 
\nonumber
\ena
The color index is denoted by $a$ and 
$C = \gamma^0\gamma^2$ is the charge 
conjugation matrix. In the baryon case we take the $u$ and $d$ quarks to
be in a $[ud]$ diquark configuration antisymmetric in spin and isospin. We
emphasize, however, that we treat the $u$ and $d$ quarks as separate dynamical
entities and not as a dynamical diquark state.  
Vertex functions in momentum space are obtained from the 
Fourier transformations of the vertex functions $F_H$ in
Eqs.~(\ref{eq:lag_Lambda}) and  (\ref{eq:lag_psi}). 
In the numerical calculations we choose a simple Gaussian form 
for the vertex functions (for both mesons and baryons): 
\be
\bar\Phi_H(-P^2) = \exp(P^{\,2}/\Lambda_H^2) \,,
\label{eq:Gauss}
\en  
where $\Lambda_H$ is a size parameter describing the distribution 
of the quarks inside a given hadron $H$. 
We use the values of these parameters fixed before 
in~\cite{Ivanov:2011aa,Dubnicka:2013vm,Gutsche:2013pp}. We would like to 
stress 
that the Minkowskian momentum variable $P^{\,2}$ turns into the Euclidean form 
$-\,P^{\,2}_E$ needed for the appropriate falloff behavior of the 
correlation function~(\ref{eq:Gauss}) in the Euclidean region.
We emphasize that any choice for the correlation function $\bar\Phi_H$ is 
acceptable
as long as it falls off sufficiently fast in the ultraviolet region of
Euclidean space. The choice of a Gaussian form for $\bar\Phi_H$ has obvious
calculational advantages.

For given values of the size parameters $\Lambda_H$
the coupling constants $g_{\Lambda_{Q}}$ and $g_{V}$ are determined by 
the compositeness condition suggested by Weinberg~\cite{Weinberg:1962hj}
and Salam~\cite{Salam:1962ap} (for a review, see~\cite{Hayashi:1967hk})
and extensively used in our approach (for details, see~\cite{Efimov:1993ei}).  
The compositeness condition
implies that the renormalization constant of 
the hadron wave function is set equal to zero   
\be 
Z_H = 1 - \Sigma^\prime_H  = 0 \,,  
\label{eq:Z=0}
\en 
where $\Sigma^\prime_H$ is the on-shell derivative of the 
hadron mass function $\Sigma_H$ with respect to its momentum. The 
compositeness condition can be seen to provide for the correct charge 
normalization for a charged bound state (see e.g.\cite{Ivanov:2011aa}).

Next we discuss the calculation of the matrix element of the 
$\Lambda_b \to\, \Lambda \,+\,V$ transition. We work in the 
so-called factorization approximation in which the matrix element
for $\Lambda_b \to \Lambda\,+\,V$ factorizes into a $(b \to s)$ 
current-induced matrix element
$<\Lambda|J^{\mu}|\Lambda_{b}>$ and a $(c \to c)$ 
current-induced vacuum to vector meson matrix element $<V|J^{\mu}|0>$. 
In our approach
the $\Lambda_b \to \Lambda$ transition is described by a two-loop Feynman-type
diagram and the current-induced vacuum to vector meson transition is 
described by a one-loop Feynman-type
diagram.  The latter diagram is proportional to the 
leptonic decay constant of the vector meson denoted by $f_V$. We have
calculated $f_{J/\psi}$ before in Ref.~\cite{Ivanov:2011aa} and have found 
$f_J = 415$ MeV in almost perfect agreement with the measured value. 
In the calculation of quark-loop diagrams we use the  
set of model parameters fixed in our previous studies.  
The model parameters are the constituent quark masses $m_q$ and
the infrared cutoff parameter $\lambda$ responsible for quark confinement.
They are taken from a fit done in 
the papers~\cite{Ivanov:2011aa,Dubnicka:2013vm} 
\be
\def\arraystretch{2}
\begin{array}{ccccccc}
     m_u        &      m_s        &      m_c       &     m_b & \lambda  &
\\\hline
 \ \ 0.235\ \   &  \ \ 0.424\ \   &  \ \ 2.16\ \   &  \ \ 5.09\ \   &
\ \ 0.181\ \   & \ {\rm GeV}
\end{array}
\label{eq: fitmas}
\en
The dimensional size parameters of the $\Lambda_b$ and $\Lambda$ baryons
have been determined in \cite{Gutsche:2013pp} by a fit to the semileptonic 
decays $\Lambda_{b} \to \Lambda_{c}+ \ell^{-}\bar \nu_{\ell}$ and 
$\Lambda_{c} \to \Lambda+ \ell^{+} \nu_{\ell}$. The resulting values are
$\Lambda_{\Lambda} = 0.490$ GeV, 
and $\Lambda_{\Lambda_b} = 0.569$ GeV. 
For the size parameter of the $J/\psi$ we take $\Lambda_{J/\psi} = 1.482$ GeV
resulting from the fit in~\cite{Ivanov:2011aa}. As of yet we cannot treat 
radial excitations in our approach. We therefore take the experimental value
$f_{\psi(2S)}=286.7$ MeV for the $\psi(2S)$.

\begin{figure}[hb] 
\begin{center}
\vspace*{1.5cm}
\epsfig{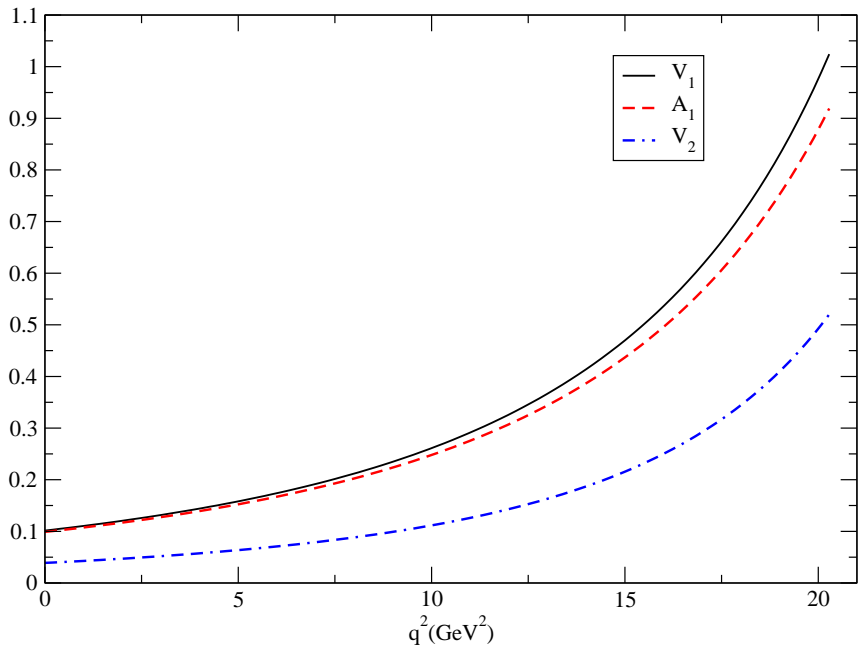}
\caption{$q^{2}$ dependence of the three form factors 
$V_1 \equiv f_{1}^{V}(q^{2})$, $A_1 \equiv f_{1}^{A}(q^{2})$ and 
$V_2 \equiv f_{2}^{V}(q^{2})$. 
The form factor $A_2 \equiv f_{2}^{A}(q^{2})$ is not shown since it would 
not be visible at the scale of the figure. 
The values of $q^{2}=0$ and $q^{2}=q^{2}_{\rm max}$ 
correspond to the maximal
and minimal recoil points, respectively.
}
\label{fig:ff}
\end{center}
\end{figure}

It should be quite clear that the evaluation of the form factors is 
technically quite
involved since it involves the calculation of a two-loop Feynman diagram
with a complex spin structure resulting from the quark propagators and the
vertex functions, which leads to a number of two-loop tensor 
integrals. To tackle this difficult
task we have automated the calculation in the form  
of FORM~\cite{Vermaseren:2000nd} and FORTRAN packages written for this purpose.
The $q^{2}$ behavior of the form factors are shown in Fig.~\ref{fig:ff}.

The results of our numerical calculations are well represented
by a double--pole parametrization 
\be\label{DPP} 
f(\hat s)=\frac{f(0)}{1 - a \hat s + b \hat{s}^2}\,, 
\en
where $\hat s=q^2/M_{\Lambda_b}^2$.   
Using such a parametrization facilitates further treatment such as the 
$q^{2}$ integrations without
having to do a numerical evaluation for each $q^{2}$ value separately.
The values of $f(0)$, $a$ and $b$ are listed  
in Table~\ref{tab:fflbs}.  
It is quite noteworthy that the numerical values
of $a$ and $b$ for each form factor in Table~\ref{tab:fflbs} are 
approximately related by $\sqrt{b}\approx a/2$ such
that the ensuing form factors are of approximate dipole form. 
The relevant scale of the effective dipole form factors is determined by 
$m_{dipole}=m_{\Lambda_{b}}/\sqrt{r}$ where $r$ is taken to be the average of
$\sqrt{b}$ and $a/2$; i.e. we take $r=(\sqrt{b}+a/2)/2$. 
The corresponding mass scale $m_{dipole}$ of the effective dipole form factor
is then given by $m_{dipole}=m_{\Lambda_{b}}/r$. One calculates  
$m_{dipole}=5.23,\,5.08,\,5.28$ and $4.32$
GeV for the four form factors in Table~\ref{tab:fflbs}. It is quite
gratifying that for each case the effective dipole masses come out to be 
close to the
average weighted mass $(m_{B_{s}}+3m_{B^{\ast}_{s}}/4)=5.31$ GeV
of the ground state $(b\bar s)$ mesons, which would set the scale for the
$q^{2}$ behavior of the form factors in a 
generalized vector dominance picture.    

In Table~\ref{tab:MR} we list our form factor results for three different
values of $q^{2}$ and compare them to the results of the light-front
diquark model calculation
of \cite{Wei:2009np} and the potential quark model calculation of 
\cite{Mott:2011cx}. At $q^2=m^2_{J/\psi}$ all three model
form factors agree for the large form factors $f^{V}_{1}$ and $f^{A}_{1}$,
while the small form factors $f^{V}_{2}$ and 
$f^{A}_{2}$ of \cite{Wei:2009np} differ from those of the other two models.  
There are larger discrepancies of the three sets of form factors for the other
two $q^{2}$ values.
In particular, at $q^{2}_{\rm max}$ the form factor values 
of~\cite{Wei:2009np} are much smaller than those of the other two models,
while at $q^{2}=0$ the large form factors $f^{V}_{1}$ and $f^{A}_{1}$
of \cite{Mott:2011cx} come out much smaller than in the two other models.
The $q^{2}=0$ values of our form factors in Table~\ref{tab:fflbs} slightly 
differ from those in Table~\ref{tab:MR} because the former are fit results
while the latter are full model results. 
The values of the form factors at $q^2=m^2_{J/\psi}$ show that
the effective interaction of the $\Lambda_b \to \Lambda$ transition is
very close to a $(V-A)$ form in all three models.

\begin{table}[hb]
\caption{Parameters for the approximated form factors
in Eq.~(\ref{DPP}) in $\Lambda_b \to \Lambda$ transitions.} 
\begin{center}
\def\arraystretch{1}
\begin{tabular}{ccccc}
\hline
       &\qquad $f_1^V$ \qquad & \qquad $f_2^V$ \qquad & 
        \qquad $f_1^A$ \qquad & \qquad $f_2^A$ \qquad \\
\hline
$f(0)$ & 0.107  & 0.043   & 0.104   & $-0.003$ \\
$a$    & 2.271  & 2.411   & 2.232   & 2.955 \\
$b$    & 1.367  & 1.531   & 1.328   & 3.620 \\
\hline
\end{tabular}
\label{tab:fflbs}
\end{center}

\vspace*{.1cm} 

\caption{Comparison of our form factor values at
$q^{2}=0$, $q^2=m^2_{J/\psi}$ and $q^2=q^2_{\rm max}$
with those obtained in \cite{Wei:2009np,Mott:2011cx}.
}
\begin{center}
\def\arraystretch{1}
\begin{tabular}{cccccccc}
\hline
  &  &  \qquad $f_1^V$ \qquad
     &  \qquad $f_2^V$ \qquad
     &  \qquad $f_3^V$ \qquad
     &  \qquad $f_1^A$ \qquad
     &  \qquad $f_2^A$ \qquad
     &  \qquad $f_3^A$ \qquad
\\
\hline
$q^2=0$ & \cite{Wei:2009np}  \qquad &
         \quad   0.1081 & 0.0311 & & 0.1065 &  0.0064 & \\
        & \cite{Mott:2011cx} \qquad &
         \quad   0.025 & 0.017  & $-\,0.0053$ & 0.028 &  0.0049 & $-\,0.019$\\
        & our  \qquad& \quad 0.10  & 0.039  & $-\,0.0017$ & 0.099 & 0.0036 
        & $-\,0.047$ \\

\hline
$q^2=m^2_{J/\psi}$ & \cite{Wei:2009np}  \qquad &
         \quad   0.248 & 0.105 & & 0.249 &  0.0214 &  \\
                  & \cite{Mott:2011cx} \qquad &
           \quad 0.255  & 0.100  & $-\,0.044$ & 0.237 &  0.020 & $-\,0.136$ \\
& our \qquad & \quad 0.25  & 0.11 & $-\,0.0097$ & 0.24 & $-\,0.0066$ &
$-\,0.13$ \\
\hline
$q^2=q^2_{\rm max}$ & \cite{Wei:2009np}  \qquad &
         \quad   0.532 & 0.204 & & 0.613 &  0.0471 &  \\
        & \cite{Mott:2011cx} \qquad &
        \quad    0.903 & 0.256 & 0.054 & 0.869 & $-\,0.072$ & $-\,0.308$ \\
        & our \qquad & \quad 1.02 
        & 0.52& $-\,0.099$ & 0.92 & $-\,0.0018$ & $-\,0.67$ \\
\hline
\end{tabular}
\label{tab:MR}
\end{center}
\end{table} 

At maximum recoil $q^{2}=0$ we can compare our results with the 
light-cone sum rules (LCSR) results of \cite{Khodjamirian:2011jp} 
on the $\Lambda_{b}\to p$ transition form factors if we assume $SU(3)$ 
to hold. In the limit of $SU(3)$ the $\Lambda_{b}\to\Lambda$ and
$\Lambda_{b}\to p$ form factors are related by 
$F(\Lambda_{b}\to\Lambda)=\sqrt{2/3} \,F(\Lambda_{b}\to p)$. This can be 
seen by using the $\bar 3 \otimes 3\,\to 8$ Clebsch-Gordan table listed in 
\cite{Kaeding:1995vq}. Based on the observation that the $[ud]$ diquark
is the $(Y=2/3,I=0)$ member of the $\bar 3$ multiplet one needs the 
Clebsch-Gordan (C.G.) coefficients 
\begin{eqnarray}
\label{clebsch}
\mbox{$\Lambda_{b}\to\Lambda$}&:&
\qquad <\mbox{\boldmath $\bar 3$},\tfrac 2 3,0,0;\,\mbox{\boldmath $3$} ,
-\tfrac 2 3,0,0|\,\mbox{\boldmath $8$},0,0,0>\,=\sqrt{2/3}\, , \\
\mbox{$\Lambda_{b}\to p$}&:&
\qquad <\mbox{\boldmath $\bar 3$},\tfrac 2 3,0,0;\, 
\mbox{\boldmath $3$},\tfrac 1 3,
\tfrac 1 2,\tfrac 1 2|\,
\mbox{\boldmath $8$},1,\tfrac 1 2,\tfrac 1 2>\,\,\,=1\,.
\nonumber
\end{eqnarray}
The labeling in~(\ref{clebsch}) proceeds according to the sequence 
$|\,\mbox{\boldmath $R$},Y,I,I_{z}>$
where $\mbox{\boldmath $R$}$ denotes the relevant $SU(3)$ representation.

The LCSR results of \cite{Khodjamirian:2011jp} have been summarized in
Table 1 of \cite{Mannel:2011xg}. We take central values of the results
listed in \cite{Mannel:2011xg} and average over the two options of 
$\Lambda_{b}$ currents. We finally multiply these numbers by $\sqrt{2/3}$ and
obtain $f^{V}_{1}=0.11\,(0.10)$, $f^{V}_{2}=0.041\,(0.039)$,
$f^{A}_{1}=0.11\,(0.099)$ and $f^{A}_{2}=0.018\,(-0.0036)$ where we have
added our model predictions in brackets. The agreement is quite
satisfactory except for the small form factor $f^{A}_{2}$ where we obtain a 
smaller value that differs in sign from that in \cite{Mannel:2011xg}.

%%%%%%%%%%%%%%%%%%%%%%%%%%%%%%%%%%%%%%%%%%%%%%%%%%%%%%%%%%%%%%%%%%%%%%%%%%%%%

\section{The heavy quark limit for the \boldmath{ $\Lambda_b\to\Lambda$} 
form factors}

%%%%%%%%%%%%%%%%%%%%%%%%%%%%%%%%%%%%%%%%%%%%%%%%%%%%%%%%%%%%%%%%%%%%%%%%%%%%

It is instructive to explore the HQL for the heavy-to-light
transition $\Lambda_b\to\Lambda$ in our form factor expressions. The HQL
corresponds to the limit $m_{\Lambda_b},m_b\to\infty $ while keeping 
the difference $m_{\Lambda_b}-m_b=\bar \Lambda $ and the size parameter
$\Lambda_{\Lambda_{b}}$ fixed. The limit has to be 
taken in the relevant expressions for the coupling constants and the form 
factors.  

First consider the local $b-$quark propagator that reduces to the static form
\be
S_b(k_1+p_1)=\frac{1}{m_b-\!\not\! k_1 - \!\not\! p_1 }\to 
\frac{1+\not\! v_1}{-2k_1v_1-2\bar\Lambda} 
+ {\cal O}\left(\frac{1}{m_b}\right),
\label{eq:prop_HQL}
\en
in the HQL. In Eq.~(\ref{eq:prop_HQL}) $p_1$ and $v_1=p_1/m_{\Lambda_b}$
denote the momentum and the four-velocity of the  $\Lambda_b$. 
The momentum $k_1$ is the loop momentum running through the loop involving 
the $b \to s$ transition. The value of the parameter 
$\bar\Lambda=m_{\Lambda_b}-m_b$ is fixed by our overall fit value for the 
$b$--quark mass [see Eq.~(\ref{eq: fitmas})]. 

Next consider the $b$-quark mass dependence of the
vertex function $F_{\Lambda_Q}(x;x_1,x_2,x_3)$ in Eq.~(\ref{eq:lag_Lambda}).
In our model the vertex function reads 
\be
F_{\Lambda_{Q}}(x;x_1,x_2,x_3) \, = \, 
\delta^{(4)}(x - \sum\limits_{i=1}^3 w_i x_i) \;  
\Phi_{\Lambda_{Q}}\biggl(\sum_{i<j}( x_i - x_j )^2 \biggr) \,, 
\label{eq:vertex}
\en 
where $\Phi_{\Lambda_{Q}}$ is the correlation function of the three 
constituent 
quarks with the coordinates $x_1$, $x_2$, $x_3$ and the masses $m_1$, $m_2$, 
$m_3$, respectively. The variable $w_i$ is defined by 
$w_i=m_i/(m_1+m_2+m_3)$ such that $\sum_{i=1}^3 w_i=1.$
In the present application $m_1=m_b$, $m_2=m_u$ and $m_3=m_d$.
In the limit $m_1\to\infty$ one has
\be
w_1 = 1-\frac{m_2+m_3}{m_1} 
+ {\cal O}\left(\frac{1}{m_1^2}\right)
\,, \qquad 
w_2 = \frac{m_2}{m_1} + {\cal O}\left(\frac{1}{m_1^2}\right)\,, \qquad 
w_3 = \frac{m_3}{m_1} + {\cal O}\left(\frac{1}{m_1^2}\right)\,.
\en
As it turns out, the next-to-leading order corrections $\frac{m_2}{m_1}$
and  $\frac{m_3}{m_1}$ contribute significantly to the HQL and cannot be 
neglected in the numerical calculations. On the other hand, we must keep 
such terms because the terms $p_{1}w_2 \sim m_2$ and $p_{1}w_3 \sim m_3$ 
occurring in the vertex function do not vanish in the heavy quark limit 
(where $p_{1}$ is the momenta of the $\Lambda_b$ baryon).   

In the HQL the coupling constant $g_{\Lambda_{Q}}$ does not depend
on the $b-$quark mass and the $\Lambda_b$ mass [there is a dependence on 
the ${\cal O}(m_Q^0)$ parameters --- $\bar\Lambda$ and $\Lambda$].   
This is specific to the three-quark system. For example, in the meson case
the meson-quark coupling constant scales as
$\sqrt{m_b}$ when $m_b\to\infty$. The constancy of $g_{\Lambda_{Q}}$ in the
HQL will be used to demonstrate the validity of the HQL for the transition 
form factors. First, we stress that the Ward identity relating 
the derivative of the mass operator and the electromagnetic vertex function 
at $p_{1} = p_{2}$ for the heavy baryon with charge $\pm 1$ is still valid. 
To show the validity of the Ward identity we consider the charged 
baryon $\Lambda_c$ (the heavy quark symmetry partner 
of the $\Lambda_b$ baryon), which has the charge $e_{\Lambda_c} = 1$. 
Using a Ward identity one can rewrite the compositeness 
condition $Z_{\Lambda_{c}}=0$ for the heavy $\Lambda_{c}$ baryon,  
in the form (see Ref.~\cite{Gutsche:2013pp})
\be
\bar u_{\Lambda_c}(p) 
\Lambda^\mu_{\Lambda_{c}}(p,p) u_{\Lambda_c}(p) = 
\bar u_{\Lambda_c}(p) \gamma^\mu u_{\Lambda_c}(p) \,,\qquad 
\not\! p \, u_{\Lambda_c}(p) = m_{\Lambda_{c}} u_{\Lambda_c}(p) \,,
\label{eq:Z=0-lam}
\en
where the electromagnetic vertex function $\Lambda^\mu_{\Lambda_{c}}(p,p)$
obtains contributions from the electromagnetic current coupling to
the quark lines (triangle contributions) and the vertices 
(bubble contributions) (see Ref.~\cite{Gutsche:2013pp} for details).
To make the HQL more transparent we perform a shift of the loop momenta
$k_1\to k_1 + (w_2+w_3)p$ and $k_2\to k_2 + w_2 p$.
One has
\bea
\Lambda^\mu_{\Lambda_{c}}(p,p) &=&
6\,g_{\Lambda_{c}}^2\,\Bla\Bla e_1 A_1^\mu -e_2 A_2^\mu + e_3 A_3^\mu\Bra\Bra
\nn
&+& 8\,g_{\Lambda_{c}}^2\,s_{\Lambda_{c}}
\Bla\Bla\Big[ Q_1 (k_1^\mu+(w_2+w_3)p^\mu)
             +Q_2(k_2^\mu+w_2 p^\mu)\Big]A_0\Bra\Bra \,, 
\label{eq:em-vert}
\ena
where the double bracket notation $<<\ldots>>$ stands for the two integrations 
over the loop 
momenta (see Ref.~\cite{Gutsche:2013pp} ). 

Also we use the definitions 
\bea
Q_1 &=& e_1 (w_2+2w_3) - e_2(w_1-w_3)  - e_3 (2w_1+w_2)\,, 
\nn
Q_2 &=&   e_1 (w_2-w_3)  - e_2(w_1+2w_3) + e_3 (w_1+2w_2)\,,  
\nn
&&
e_1 \equiv e_c = \tfrac23, \quad e_2 \equiv e_u =\tfrac23, \quad 
e_3 \equiv e_d = - \tfrac13, 
\label{Q1Q2}
\ena
and
\bea
A_0 &=& \bar\Phi_{\Lambda_{Q}}^2(-z_0)
S_1(k_1+p)
\Tr\left[S_2(k_2)\gamma^5 S_3(k_2-k_1)\gamma^5\right]\,,
\nn 
A^\mu_1&=& \bar\Phi^2_{\Lambda_{Q}}(-z_0)
S_1(k_1+p)\gamma^\mu S_1(k_1+p)
\Tr\left[S_2(k_2)\gamma^5 S_3(k_2-k_1)\gamma^5\right]\,, 
\nn
A^\mu_2 &=& \bar\Phi^2_{\Lambda_{Q}}(-z_0)
S_1(k_1+p)
\Tr\left[S_2(k_2)\gamma^\mu S_2(k_2) 
\gamma^5 S_3(k_2-k_1)\gamma^5\right]\,,
\nn 
A^\mu_3 &=& \bar\Phi^2_{\Lambda_{Q}}(-z_0)
S_1(k_1+p)
\Tr\left[S_2(k_2)
\gamma^5 S_3(k_2-k_1)\gamma^\mu S_3(k_2-k_1)\gamma^5\right]\,.
\label{eq:bricks}
\ena
Here $s_{\Lambda_{Q}} \equiv 1/\Lambda_{\Lambda_{Q}}^2$ and the argument
of the vertex function is
\be
z_0 =\tfrac12(k_1-k_2+w_3 p)^2+\tfrac16(k_1+k_2+(2w_2+w_3)p)^2\,.
\label{eq:arg}
\en
The calculational techniques of the matrix elements in the heavy 
quark limit can easily be demonstrated for the example of the 
structure integral occurring in the heavy meson case 
(the extension to the heavy baryon case is straightforward) 
\be
I_2(\bar\Lambda)=\int \frac{d^4k}{i\pi^2} 
e^{s k^2}\, S_Q(k_1+p) \, \frac{1}{m^2-k^2}\,,
\label{eq:ex1}
\en
where $s=1/\Lambda^2$. 
For simplicity we only keep the product of the heavy quark propagator 
and the denominator of the light quark propagator. 
We start with  the Schwinger representation for the quark propagators
assuming that both loop and external momenta are Euclidean.
In the Euclidean region the denominator of the quark propagator
is positive and the integral over the Schwinger parameter is absolutely
convergent. However, to use the Schwinger representation  
for the heavy quark propagator in Eq.~(\ref{eq:prop_HQL}) in a straightforward
way is not quite correct because the HQL has to be taken in
Minkowski space where the denominator is not necessarily positive.
We will use the heavy quark propagator in the form    
\be
S_Q(k_1+p)= 
m_Q(1+\not\! v)\int\limits_0^\infty d\alpha e^{-\alpha(m_Q^2-(k_1+p)^2)}\, ,
\label{eq:prop_HQL_1}
\en
assuming again that all momenta in the exponential are in
the Euclidean region. 
For the numerator of the heavy quark propagator we take the HQL: 
$m_Q  \, + \not\! k \, + \not\! p \to m_Q ( 1+\not\! v)$. 
Next we demonstrate how to proceed with the HQL for such a representation. 
As mentioned above, we start 
in the Euclidean region where $k^2\le 0$ and $(k+p)^2\le 0$.
By using Schwinger's representation for the heavy quark propagator and 
the denominator of the light quark propagator, 
scaling the Schwinger parameters $\alpha_i\to t\alpha_i$
and imposing an infrared cutoff, we arrive at
\be
I_2(\bar\Lambda)=m_Q\,(1+\not\! v)\,
\int\limits_0^{1/\lambda^2} dt\frac{t}{(s+t)^2}
\int\limits_0^1 d\alpha e^{-t(\alpha m_Q^2+(1-\alpha)m^2-\alpha(1-\alpha)p^2)
+\frac{st}{s+t}\alpha^2p^2} .
\label{eq:ex2}
\en
We will use this representation for the analytical continuation
to the physical region $p^2=(m_Q+\bar\Lambda)^2$ with $m_Q\to \infty$.
Note that in a theory without a cutoff ($\lambda\to 0$)
the integral $I_2(\bar\Lambda)$ has a branch point at $\bar\Lambda=m$.
The confinement ansatz allows one to remove this singularity.
Then we scale the integration variable $\alpha\to\alpha/m_Q$
with $m_Q\to\infty$. Finally, one has
\be
I_2(\bar\Lambda)\biggl|_{m_Q\to\infty} \longrightarrow \, 
I^{\rm HQL}_2(\bar\Lambda)
=(1+\not\! v)\,\int\limits_0^{1/\lambda^2} dt\frac{t}{(s+t)^2}
\int\limits_0^\infty d\alpha e^{-t(\alpha^2-2\alpha\bar\Lambda+m^2)
+\frac{st}{s+t}\alpha^2} .
\label{eq:ex3}
\en

The calculation of the HQL for the coupling constant $g_{\Lambda_{b}}$ and 
the transition form factors $\Lambda^\mu_{\Lambda_b \to \Lambda}(p_1,p_2)$ 
proceed in the described way. All analytical calculations are done 
by FORM~\cite{Vermaseren:2000nd} and the numerical calculations are done 
using FORTRAN. One finds
\bea
g_{\Lambda_b} = \left\{
\begin{array}{ll}
65.23\,\,\, \text{GeV}^{-2}  & \qquad  \text{exact}\,, \\
59.44\,\,\,  \text{GeV}^{-2} & \qquad \text{HQL} \,. 
\end{array} \right.
\label{coupling}
\ena 
The coupling constant in the HQL is smaller than the exact coupling constant 
only by about $\,10\,\%$ which shows that one is quite close to the
HQL for the $\Lambda_{b} \to \Lambda$ transitions. 
When calculating the HQL for the coupling constant it is important
to keep the numerical value of the parameter $\bar\Lambda_{b}$ fixed
at its physical value $\bar\Lambda_{b}=m_{\Lambda_{b}}-m_{b}=0.53$ GeV.
In fact, the value of $g_{\Lambda_b}^{\rm HQL}$ depends very sensitively
on the choice of the parameter $\bar\Lambda_{b}$. For example, if one puts  
$\bar\Lambda_{b}=0$ then 
one calculates $g_{\Lambda_b}^{\rm HQL}=185.36$ GeV$^{-2}$ which differs
significantly from the exact result $g_{\Lambda_b}=65.23$ GeV$^{-2}$ .
This demonstrates how important it is to keep the physical value
of $\bar\Lambda_{b}$. 
It is interesting to 
compare our results for the coupling constants $g_{\Lambda_b}$ and 
$g_{\Lambda_c}$. For $g_{\Lambda_c}$ one finds 
\bea 
g_{\Lambda_c} = \left\{
\begin{array}{ll}
69.88\,\,\, \text{GeV}^{-2}  & \qquad  \text{exact}\,, \\
60.01\,\,\,  \text{GeV}^{-2} & \qquad \text{HQL}\,. 
\end{array} \right.
\label{couplingc}
\ena 
One can see that there is a small difference between the coupling
constants in the exact case, 
while in the HQL they are practically degenerate. This happens because, as
was emphasized 
before, the coupling constant $g_{\Lambda_Q}$ in the HQL does not depend 
on the heavy quark mass. A small difference of the coupling constants 
$g_{\Lambda_b}$ and $g_{\Lambda_c}$ in HQL is due to 
dependence on ${\cal O}(m_Q^0)$ parameters --- $\bar\Lambda$ and $\Lambda$). 
  
The vertex function $\Lambda^\mu_{\Lambda_b \to \Lambda}(p_1,p_2)$ that 
describes the heavy-to-light $\Lambda_b\to\Lambda$ transition reads
\bea
\Lambda^\mu_{\Lambda_b\to\Lambda}(p_1,p_2) &=&
6\,g_{\Lambda_b}\,g_{\Lambda}\,
\Bla\Bla\bar\Phi_{\Lambda}(-z_s) \bar\Phi_{\Lambda_b}(-z_b)
\nonumber\\
&\times&
S_s(k_1+p_2)\Gamma^\mu S_b(k_1+p_1)\,
\Tr\left[S_u(k_2)\gamma^5 S_d(k_2-k_1)\gamma^5\right]\Bra\Bra \,,
\label{eq:LbLs}
\ena 
where 
\bea 
z_i &=&  \tfrac12(k_1-k_2+w_3^i\, p_2)^2 
       + \tfrac16(k_1+k_2+(2w_2^i+w_3^i)\,p_2)^2\,. 
\ena
In the present application we have to consider
two cases of $\Gamma^\mu$: $\gamma^\mu$ and  
$\gamma^\mu\gamma^5$. 
The loop calculation contains the variable $v_1p_2$ that, for a given 
$q^{2}$, is fixed by the kinematics of the process through
\be
v_1p_2 = \frac{m_{\Lambda_b}}{2}\left(1+\frac{m_{\Lambda}^2-q^2}
{m_{\Lambda_b}^{2}}\right)\,, 
\label{eq:kinematics}
\en
where $p_2$ and $m_{\Lambda}$ are the momentum and mass of the  $\Lambda$. 

\begin{figure}[ht] 
\begin{center}
\epsfig{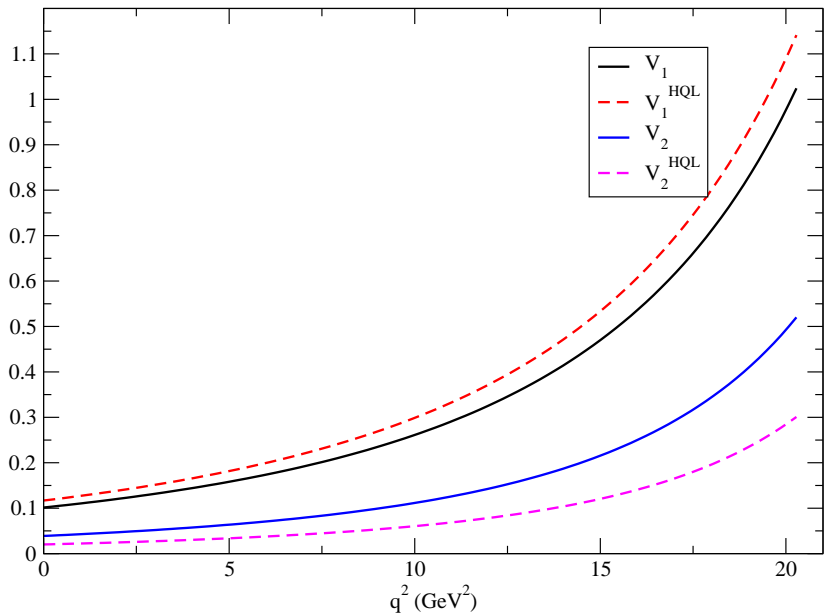} 
\caption{Comparison of the form factors 
$V_i \equiv f_{i}^{V}(q^{2})$ vs. $V_i^{\rm HQL} \equiv 
f_{i}^{V,\,{\rm HQL}}(q^{2}) \ (i=1,2)$. 
}
\label{fig:HQL}
\end{center}
\end{figure}

In the heavy quark limit we reproduce the form factor structure 
derived previously from heavy quark effective theory 
\cite{Hussain:1990uu,Hussain:1992rb,Mannel:1990vg}, which is usually written
in the form
\be
\bar u_2(p_2)\Lambda^\mu_{\Lambda_b\to\Lambda}(p_1,p_2) u_1(p_1)
=
\bar u_2(p_2)
\Big[ F_1(q^2)+ F_2(q^2)\not\! v_1 \Big]\Gamma^{\mu} u_1(p_1).
\label{eq:hqet}
\en
Again, for the present application, $\Gamma^{\mu}$ is $\gamma^{\mu}$ or
$\gamma^\mu\gamma^5$.
From Eq.~(\ref{eq:hqet}) one finds
\be
f_{1}^{V,\,{\rm HQL}}=f_{1}^{A,\,{\rm HQL}} 
= F_{1}+\frac{M_{2}}{M_{1}}F_{2}\,,\qquad
f_{2}^{V,\,{\rm HQL}}=f_{2}^{A,\,{\rm HQL}}=-F_{2}\,,\qquad
f_{3}^{V,\,{\rm HQL}}=f_{3}^{A,\,{\rm HQL}}= F_{2}\,,
\label{eq:HQL_relations}
\en
i.e. there are only two independent form factors in the HQL. We emphasize 
that the form factor relations
Eq.~(\ref{eq:HQL_relations}) are valid  in the full kinematical region
$4m_{\ell}^{2} \le q^{2} \le (M_{\Lambda_{b}}- M_{\Lambda})^{2}$.
In Fig.~\ref{fig:HQL} we plot the $q^{2}$ dependence of the form factors 
$f^{V}_{1}$ and $f^{V}_{2}$ and compare them to the corresponding form 
factors calculated in the HQL. For the large form factor $f^{V}_{1}$ the
HQL form factor exceeds the full form factor by $O(10\,\%)$ while the
small form factor $f^{V}_{2}$ is lowered by $O(50\,\%)$.

It is interesting to compare our HQL form factors with the corresponding
static lattice results presented in \cite{Detmold:2013nia}. Although
the main concern of \cite{Detmold:2013nia} was the $\Lambda_{b} \to p$ form 
factors, the authors also present results on the $\Lambda_{b} \to \Lambda$
form factors in their Fig.~\ref{fig:HQL}. 
At $q^{2}_{\rm max}$, where the lattice calculations are most reliable, 
one reads off from their Fig.~\ref{fig:HQL} 
$F_{1}=1.28\pm0.05$ and $F_{2}=-0.30 \pm 0.02$, where the errors are only 
statistical, compared to the HQL limiting 
values $F_{1}=1.21$ and $F_{2}=-0.30$ in our model. The agreement is
satisfactory. As concerns the $q^{2}$ behavior of the form factors our HQL 
form factors fall off somewhat more steeply than the static lattice form 
factors. At $q^{2}=12.4\,{\rm GeV^{2}}$, where the lattice results may 
not be so reliable, one finds from Fig.~\ref{fig:HQL} 
of~\cite{Detmold:2013nia} 
$F_{1}=0.64\pm0.1$ and $F_{2}=-0.11 \pm 0.04$, where the errors
now include the systematic errors, compared to
our values  $F_{1}=0.31$ and $F_{2}=-0.07$.  
We mention that the static results given in \cite{Detmold:2013nia}  
approximately satisfy the $SU(3)$ relation
$F_{i}(\Lambda_{b}\to\Lambda)=\sqrt{2/3} \,F_{i}(\Lambda_{b}\to p)\,\,(i=1,2)$
derived in Eq.~(\ref{clebsch}). 
 
At the large recoil end of the $q^{2}$ spectrum soft collinear effective
theory predicts that the form factor $F_{2}$ vanishes in the heavy 
quark limit at the leading order of $\alpha_{s}$ 
\cite{Mannel:2011xg,Feldmann:2011xf}. The plot of
$f_{2}^{V,\,{\rm HQL}}(q^{2})=-F_{2}(q^{2})$ in Fig.~\ref{fig:HQL}   
shows that our prediction for $F_{2}$ reaches a very small 
albeit nonzero value at large recoil.

%%%%%%%%%%%%%%%%%%%%%%%%%%%%%%%%%%%%%%%%%%%%%%%%%%%%%%%%%%%%%%%%%%%%%%%%%%%%%%

\section{Numerical results} 

%%%%%%%%%%%%%%%%%%%%%%%%%%%%%%%%%%%%%%%%%%%%%%%%%%%%%%%%%%%%%%%%%%%%%%%%%%%%%%

We begin by presenting our numerical results for the decay
$\Lambda_{b}\to \Lambda\,+\,J/\psi(\to e^{+}e^{-},\mu^{+}\mu^{-})$ 
for which one can safely use the zero lepton
mass approximation (see Table~\ref{tab:velocity}).
Our results are presented in Tables~\ref{tab:rates}-\ref{tab:asym_hel} 
where we compare them with the available data~\cite{Aaij:2013oxa,pdg12} 
and predictions of other theoretical 
approaches~\cite{Cheng:1995fe}-\cite{Mott:2011cx}. In Table V we present
our result on the branching fraction $B(\Lambda_{b}\to\Lambda J/\psi)$ and 
compare
it with data and the results of other theoretical models. The data value
of $B(\Lambda_{b}\to\Lambda J/\psi)=(5.8 \pm 0.8) \times 10^{-4}$ is based on 
the PDG13 value for 
$\Gamma(\Lambda_{b}\to\Lambda J/\psi)\times B(b \to \Lambda_{b})$
given in~\cite{pdg12} and a value of $B(b \to \Lambda_{b})=0.1$ as used
in previous editions of the PDG. Our result on the branching fraction
is based on the lifetime measurement $1.429\times 10^{-12}s$ as listed
in the 2013 update of the PDG~\cite{pdg12}. If one would instead take the value
of $1.482\times 10^{-12}s$ reported in~\cite{Aaij:2013oha}, one would have 
to scale our result on 
the branching fraction upward by $3.7\,\%$. For easy comparison we have 
taken the freedom to present the results of~\cite{Wei:2009np,Mott:2011cx}
using our parameters ($\Lambda_{b}$ life time, CKM matrix elements and Wilson 
coefficients). Our branching fraction is slightly larger than those
of~\cite{Wei:2009np,Mott:2011cx}. All three branching fractions are somewhat 
larger than the experimental PDG average value. To judge on the significance
of this discrepancy one would have to wait for an absolute measurement
of the branching fraction $B(\Lambda_{b}\to\Lambda J/\psi)$. We mention
that the remaining theoretical branching fractions in 
Table~\ref{tab:rates} would have
to be readjusted upward by $(4-7)\%$ (depending on the year of publication) 
if the new 2013 lifetime measurement of 
the LHCb Collaboration~\cite{Aaij:2013oha} is used. 

In Table VI we present
our result for the asymmetry parameter $\alpha_{b}$ and compare
it with the data and the results of other theoretical models. In agreement
with the measurement the theoretical results on the asymmetry parameter come 
out to be quite small. Since the measurement carries large error bars,     
one cannot really draw any conclusions on the quality of the agreement
between the experiment and the model predictions.  

In Table VII we compare our predictions for the asymmetry parameters and
the moduli squared of the normalized helicity amplitudes with the
corresponding measurements of~\cite{Aaij:2013oxa}. The fourth column of 
Table~\ref{tab:asym_hel} contains our predictions for the corresponding
quantities in the decay $\Lambda_{b}\to\Lambda\,+\,\psi(2S)$, which differ 
notably from the corresponding quantities in the 
$\Lambda_{b}\to \Lambda\,+\,J/\psi$ mode.
The numerical results clearly show the dominance of the 
$\lambda_{\Lambda}=-1/2$ helicity configurations in both cases where the
dominance is more pronounced for the $\Lambda_{b}\to \Lambda J/\psi$ mode. 
For the $\Lambda_{b}\to \Lambda J/\psi$ mode the agreement between
our results and the data is quite satisfactory. 

Using our results in Table~\ref{tab:asym_hel} we can calculate the
three measures Eqs.~(\ref{afbtheta})-(\ref{convex}) characterizing the single 
angle decay distributions 
Eqs.~(\ref{wtheta}-\ref{wtheta2}). For the polarization related 
FB asymmetry $A_{FB}\,\big|_{\theta}$ calculated in 
Eq.~(\ref{afbtheta}) 
one obtains
\begin{equation}
A_{FB}\,\big|_{\theta}=-0.035\, P_{b} \,. 
\label{afbtheta_fin}
\end{equation}
The analyzing power related to the measurement of the 
$\Lambda_{b}$--polarization $P_{b}$ is quite
small due to the fact that in our model 
$|\widehat H_{+\tfrac12 0}|^2 \approx |\widehat H_{+\tfrac12 +1}|^2 \approx 0$
and $|\widehat H_{-\tfrac12 0}|^2 \approx |\widehat H_{-\tfrac12 -1}|^2$
leading to a very small value of the parameter $\alpha_{b}$. 
This is reflected in the poor precision of the polarization measurement 
$P_{b}=0.05\pm0.07\pm0.02$ reported 
by the LHCb Collaboration~\cite{Aaij:2013oxa}. 

The second measure $A_{FB}\,\big|_{\theta_{1}}$ related to the hadron-side
decay $\Lambda\to p\pi^{-}$ is nearly maximal in our model due to the fact
that the longitudinal polarization of the daughter baryon $\Lambda$ is close
to its maximal value. From Table~\ref{tab:asym_hel} one finds 
$P^{\ell}_{\Lambda}=2r_{1}-\alpha_{b}=-0,99$ and $P^{\ell}_{\Lambda}=-0.97$
for the decays $\Lambda_{b}\to \Lambda\,+\,J/\psi$ and 
$\Lambda_{b}\to \Lambda\,+\,\psi(2S)$, respectively. For the decay
$\Lambda_{b}\to \Lambda\,+\,J/\psi$ the FB asymmetry is given by 
[see Eq.~(\ref{afbtheta1})] 
\begin{equation}
A_{FB}\,\big|_{\theta_{1}}=P^{\ell}_{\Lambda}\,\alpha_{\Lambda}
=(2r_{1}-\alpha_{b})\,\alpha_{\Lambda}=-0.64\,(-0.62) \,, 
\label{afbtheta1_fin}
\end{equation}
where we have used the experimental value for the asymmetry parameter 
$\alpha_{\Lambda}=0.642$ \cite{pdg12} and where we have added the
corresponding number for the $\Lambda_{b}\to\Lambda\,\psi(2S)$ mode in round 
brackets. 

\begin{table}[hb]
\begin{center}
\caption{Branching ratio $B(\Lambda_b \to \Lambda\,+\,J/\psi)$ 
(in units of $10^{-4}$).} 

\vspace*{.1cm}

\def\arraystretch{1}
    \begin{tabular}{|c|c|c|}
      \hline
Our result  & Theoretical predictions & Data~\cite{pdg12} \\
\hline
8.9 & 2.1~\cite{Cheng:1995fe}; 
      1.6~\cite{Cheng:1996cs}; 
      2.7~\cite{Ivanov:1997ra}; 
      6.04~\cite{Fayyazuddin:1998ap};   
    & $5.8 \pm 0.8$~(PDG average)~\cite{pdg12}\\
    & 2.49~\cite{Mohanta:1998iu}; 
      $3.45 \pm 1.81$~\cite{Chou:2001bn}; 
      8.4~\cite{Wei:2009np}; 8.2~\cite{Mott:2011cx} 
    &\\
\hline
\end{tabular}
\label{tab:rates}
\end{center}

\begin{center}
\caption{Asymmetry parameter $\alpha_b$.} 

\vspace*{.1cm}

\def\arraystretch{1}
    \begin{tabular}{|c|c|c|}
      \hline
Our result  & Theoretical predictions & Data~\cite{Aaij:2013oxa} \\
\hline
$-0.07$ & $-0.11$~\cite{Cheng:1995fe}; 
      $-0.10$~\cite{Cheng:1996cs}; 
      $-0.21$~\cite{Ivanov:1997ra}; 
      $-0.18$~\cite{Fayyazuddin:1998ap};   
    & $-0.04 \pm 0.17 \pm 0.07$ \\
    & $-0.208$~\cite{Mohanta:1998iu}; 
      $- 0.155 \pm 0.015$~\cite{Chou:2001bn}; 
      $-0.10$~\cite{Wei:2009np}, $-0.09$~\cite{Mott:2011cx}  
   & \\
\hline
\end{tabular}
\label{tab:asym}
\end{center}

\begin{center}
\caption{Asymmetry parameters and moduli squared of normalized
helicity amplitudes.} 

\vspace*{.1cm}

\def\arraystretch{1}
    \begin{tabular}{|c|c|c|c|}
      \hline
Quantity &  Data~\cite{Aaij:2013oxa} 
&\multicolumn{2}{c|}{Our results}\\
\cline{2-4}
& $\Lambda J/\psi$ mode  
& $\Lambda J/\psi$ mode
& $\Lambda \psi(2S)$ mode \\
\hline 
$\alpha_b$       & $ - 0.04 \pm 0.17 \pm 0.07$ & - 0.07 & 0.09 \\
\hline
$r_0$            & $ 0.57 \pm 0.02 \pm 0.01$   &  0.53  & 0.45 \\
\hline   
$r_1$            & $-0.59 \pm 0.10 \pm 0.05$   & - 0.53 & -0.44 \\
\hline
$|\hat H_{+\tfrac12 0}|^2$ 
&  $-0.01 \pm 0.04 \pm 0.03$ 
&  0.46$\times 10^{-3}$ 
&  0.33$\times 10^{-2}$ 
\\[2ex]
\hline
$|\hat H_{-\tfrac12 0}|^2$ 
&  $0.58 \pm 0.06 \pm 0.03$ 
&  0.53  
&  0.45  
\\[2ex]
\hline
$|\hat H_{-\tfrac12 -1}|^2$ 
&  $0.49 \pm 0.05 \pm 0.02$ 
&  0.47 
& 0.54 
\\[2ex]
\hline
$|\hat H_{+\tfrac12 +1}|^2$ 
&  $-0.06 \pm 0.04 \pm 0.03$ 
&  0.31$\times 10^{-2}$  
&  0.12$\times 10^{-1}$  
\\[2ex]
\hline
\end{tabular}
\label{tab:asym_hel}
\end{center}
\end{table} 

\begin{figure}[ht] 
\begin{center}
\vspace*{1cm} 
\epsfig{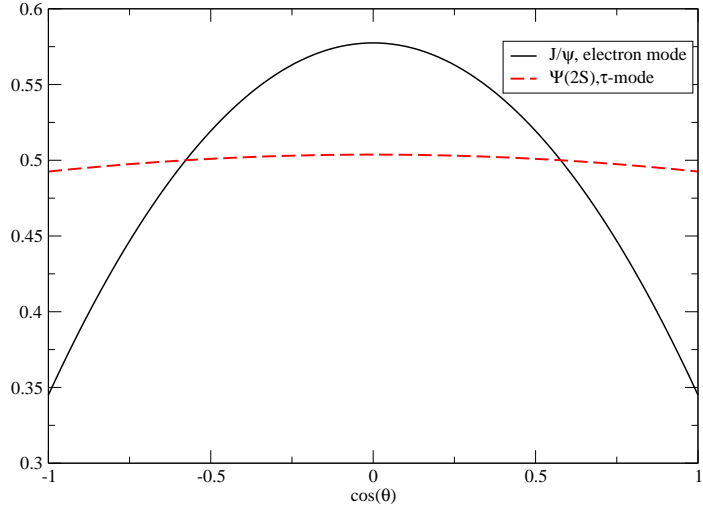} 
\caption{Polar angle distribution $\hat W(\theta_{2})$ 
for the two cases --- $\Lambda_{b} \to \Lambda\,+\,J/\psi(\to e^+e^-)$ and  
$\Lambda_{b} \to \Lambda\,+\,\psi(2S)(\to \tau^+\tau^-)$.}
\label{fig:psi_lepton}
\end{center}
\end{figure}

Next we discuss the lepton-side 
$\cos\theta_{2}$ distribution~(\ref{wtheta2}) for the decay
$\Lambda_{b}\to \Lambda \,V(\to \ell^{+}\ell^{-})$ which is governed by
the polarization of the vector charmonium state $V$. The 
transverse/longitudinal composition of the vector charmonium state
$V$ is given by $\hat U:\hat L=(1-r_{0}):r_{0}$, 
where $\hat U$ is the sum of the transverse
helicity contributions 
$|\widehat H_{+\tfrac12 +1}|^2 - |\widehat H_{-\tfrac12-1}|^2$. 
As mentioned before the electromagnetic decay $V\to\ell^+\ell^-$
is not sensitive to the difference of the two transverse helicity
contributions. Table~\ref{tab:asym_hel} shows that the transverse and 
longitudinal states are approximately
equally populated for both vector charmonium states leading to an
approximate angular decay distribution of 
$\widehat W(\theta_{2})\sim \frac{3}{16}(3-\cos^{2}\theta_{2})$ for 
$\varepsilon=0$. 
The exact numbers are $\hat U:\hat L=0.47:0.53$ and $0.55:0.45$ for
$V=J/\psi$ and $V=\psi(2S)$, respectively. In 
Fig.~\ref{fig:psi_lepton} we plot the
$\cos\theta$ distribution for the decay 
$\Lambda_{b}\to \Lambda \,J/\psi(\to e^{+}e^{-})$. The convexity of
the distribution is 
characterized by the convexity parameter $c_{f}$ defined in 
Eq.~(\ref{convex}) for which we obtain 
$c_{f}=-0.44$. Negative convexities correspond to a downward open
parabola for the decay distribution as is also evident 
from Fig.~\ref{fig:psi_lepton}. Figure~\ref{fig:psi_lepton} also contains a 
plot of the
$\cos\theta_{2}$ distribution for the case 
$\Lambda_{b}\to \Lambda\,\psi(2S)(\to\,\tau^{+}\tau^{-})$ for which the 
convexity parameter is given by $c_{f}=-0.016$ implying an almost 
flat $\cos\theta_{2}$ distribution. Because of the small factor 
$v^{2}=1-4\varepsilon=0.071$ in the
$\cos\theta_{2}$ distribution Eq.~(\ref{wtheta2}) the lepton side has lost 
all of its analyzing power in this case. 

As has already been argued after Eq.~(\ref{azimuth}) one expects a 
minimal azimuthal correlation of the two decay planes spanned by $(p\pi^{-})$
and $(\ell^{+}\ell^{-})$. In the zero lepton mass approximation one obtains 
the azimuthal decay distribution [see Eq.~(\ref{azi})]
\begin{eqnarray}
A_{FB}(\chi)&=&
\frac{\pi}{8\sqrt{2}}\,\alpha_{\Lambda}\,
\rm{Re}\big(\widehat H_{\tfrac 12 1}
\widehat H^{\dagger}_{-\tfrac 12 0}- 
\widehat H_{-\tfrac 12 -1}\widehat H^{\dagger}_{\tfrac 12 0}\big)\,\cos\chi
\nonumber\\
&=&-0.0046\,\cos\chi \,, 
\end{eqnarray}
where the numerical value can be obtained from the entries in 
Table\ref{tab:asym_hel} with the additional information that our model 
helicity amplitudes
(which are real) with helicities 
$\lambda_{\Lambda}=-1/2$ and $\lambda_{\Lambda}=1/2$ are positive and negative,
respectively.
As expected, the azimuthal correlation between the two decay
planes spanned by $(p\pi^{-})$
and $(\ell^{+}\ell^{-})$ is negligibly small. We do not write out the
result for the decay
$\Lambda_{b}\to \Lambda\,+\,\psi(2S)(\to\,\tau^{+}\tau^{-})$ 
since the additional factor of 
$(1-4\varepsilon)/(1+2\varepsilon)=0.049$ in~(\ref{wtheta2}) 
reduces the correlation measure to a value close to zero.

Finally, we calculate the cascade 
$\Lambda_b\to\Lambda\,+\,\psi(2S)(\to \ell^+\ell^-)$-decay width
by using the zero width approximation
\be
{\rm B}(\Lambda_b \to \Lambda\,+\,\psi(2S)(\to \ell^+ \ell^-) ) 
=  
{\rm B}(\Lambda_b \to \Lambda\,+\,\psi(2S)) \  
{\rm B}(\psi(2S) \to \ell^+ \ell^-) \,. 
\label{eq:ZWA}
\en 
We take the value of the leptonic decay constant
$f_{\psi(2S)} = 286.7$ MeV from the electronic mode
measured experimentally and employ the formula
\be 
\Gamma(\psi(2S) \to \ell^+  \ell^-) = 
\frac{16 \pi \alpha^2}{27} \, \frac{f_{\psi(2S)}^2}{m_{\psi(2S)}} \, 
\sqrt{1 - \frac{4 m_\ell^2}{m_{\psi(2S)}^2}}  
\, \biggl(1 + \frac{2 m_\ell^2}{m_{\psi(2S)}^2}\biggr)   
\label{eq:leptonic}
\en 
to evaluate the other modes.
The results for the branching ratios
of the decays $\Lambda_b \to \Lambda\,+\,\psi(2S)(\to \ell^+  \ell^-)$
are given in Table~\ref{tab:psi(2S)-rates}. Again,
one can see that the $\tau$-lepton mass plays an essential role
in reducing the value of the decay width as compared
to the electron and muon modes. 
Our prediction for the branching fraction of the 
$\Lambda_{b}\to \Lambda\,+\,\psi(2S)$ 
transition is (based on the lifetime value 
$\tau_{\Lambda_{b}}=1.429\times 10^{-12}s$)
\bea
B(\Lambda_b \to \Lambda\,+\,\psi(2S)) =  7.25 \times 10^{-4} \,. 
\ena

\begin{table}[htb]
\begin{center}
\caption{Branching ratios 
$B(\Lambda_b \to \Lambda\,+\,\psi(2S)(\to \ell^+  \ell^-))$ 
in units of $10^{-6}$.} 

\vspace*{.1cm}

\def\arraystretch{1}
    \begin{tabular}{|c|c|}
      \hline
Mode & Our results \\ \hline
$\Lambda_b \to \Lambda\,+\,e^+ e^-$        &  5.61 \\
\hline
$\Lambda_b \to \Lambda\,+\, \mu^+ \mu^-$    &  5.61 \\
\hline
$\Lambda_b \to \Lambda\,+\, \tau^+ \tau^-$  &  2.18 \\
\hline
\end{tabular}
\label{tab:psi(2S)-rates}
\end{center}
\end{table}

\section{Summary} 
We have performed a detailed analysis of the decay process
$\Lambda_b \to \Lambda \,+\, J/\psi $ in the 
framework of the covariant quark model.
We have worked out two variants of threefold joint angular decay 
distributions in the cascade decay 
$\Lambda_b\to \Lambda(\to p\pi^-)\,+\,J/\psi(\to\ell^+\ell^-)$ for polarized
and unpolarized $\Lambda_{b}$ decays. We have reported our numerical results 
on helicity amplitudes, on 
the rate and on the asymmetry parameters in the decay
processes $\Lambda_b \to \Lambda\,+\,J/\psi$ and 
$\Lambda_b \to \Lambda\,+\,\psi(2S)$. We have included the decay 
$\Lambda_b \to \Lambda \,+\,\psi(2S)$ in our analysis since this
decay allows one to discuss nonzero lepton mass effects in the kinematically 
allowed decay $\Lambda_b \to \Lambda\,+\,\psi(2S)(\to\tau^{+}\tau^{-})$. 
We confirm expectations from the naive quark model that the transitions into 
the $\lambda_{\Lambda}=1/2$ helicity states of the daughter baryon $\Lambda$ 
are strongly suppressed leading to a near maximal negative polarization of the
$\Lambda$. For the same reason
the azimuthal correlation between the two decay planes spanned by $(p\pi^{-})$
and $(\ell^{+}\ell^{-})$ is negligibly small.
We have compared our results with the available experimental
data and with the results of other theoretical approaches.
In a separate section we have presented form factor results
over the whole accessible range of $q^{2}$ values. These results are close 
to lattice results at minimum recoil and to LCSR results at maximum recoil.

\begin{acknowledgments}

This work was supported by the DFG under Contract No. LY 114/2-1.  
M.A.I.\ acknowledges the support from Mainz Institute 
for Theoretical Physics (MITP) and the Heisenberg-Landau Grant.  
The work was also partially supported under the project 2.3684.2011 of 
Tomsk State University. 
This work is also partially supported by the Italian Miur PRIN 2009. 

\end{acknowledgments}

\end{document}